\documentclass{sig-alternate-10pt}

\usepackage{paralist}

\usepackage[ruled,vlined]{algorithm2e}

\usepackage{url}
\usepackage{breakurl}
\usepackage{amssymb}
\usepackage{amsmath} 
\usepackage{verbatim}
\usepackage{array}
\usepackage{graphicx}
\usepackage{epstopdf}
\usepackage{stmaryrd}
\usepackage{dsfont}
\usepackage{subfig}
\usepackage{soul}

\usepackage{framed}
\usepackage[dvipsnames]{color}
\definecolor{shadecolor}{gray}{0.85}

\newif\iflong 
\longfalse

\newif\ifcomm
\commtrue

\newtheorem{property}{Property}

\renewcommand{\P}{{\mathbb P}}                         

\newenvironment{proof-sketch}{{\noindent\em Proof Sketch.}\hspace*{0.3em}}{\qed\medskip}
\newenvironment{proof-of}[1]{{\noindent\em Proof of #1.}\hspace*{0.3em}}{\qed\medskip}

\newcounter{assumption}
\newcommand{\theassumptionletter}{A}
\renewcommand{\theassumption}{\theassumptionletter\arabic{assumption}}

\newcommand{\beq}{\begin{equation}}
\newcommand{\eeq}{\end{equation}}
\newcommand{\beqa}{\begin{eqnarray}}
\newcommand{\eeqa}{\end{eqnarray}}
\newcommand{\beqan}{\begin{eqnarray*}}
\newcommand{\eeqan}{\end{eqnarray*}}
\newcommand{\ben}{\begin{eqnarray*}}
\newcommand{\een}{\end{eqnarray*}}

\ifcomm
   \newcommand\comm[1]{\textcolor{blue}{ #1}}
\else
   \newcommand\comm[1]{}
   \renewcommand{\todo}[1]{}
\fi

\newcommand{\pp}{\mbox{$\mathbf{p}$}}

\usepackage{amssymb}
\usepackage{graphicx}

\usepackage{url}
\urldef{\mailsa}\path|hanguyen4@uh.edu, rzheng@mcmaster.ca|

\begin{document}


\title{A Data-driven Study of Influences in Twitter Communities}


%
%
\numberofauthors{2}
\author{
\alignauthor
Huy Nguyen \\
\affaddr{Product Design and Development Dept.} \\
\affaddr{IHS Inc.} \\
\affaddr{8584 Katy Freeway} \\
\affaddr{Houston, Tx, USA} \\
\email{huy.nguyen@ihs.com} \\
\alignauthor
Rong Zheng \\
\affaddr{Dept. of Computing and Software} \\
\affaddr{McMaster University} \\
\affaddr{1280 Main St. West}\\
\affaddr{Hamilton, ON, Canada}\\
\email{rzheng@mcmaster.edu} \\
}
%


%
%

\maketitle

\begin{abstract}
This paper presents a quantitative study of Twitter, one of the most popular
micro-blogging services, from the perspective of user influence.  We crawl
several datasets from the most active communities on Twitter and obtain 20.5
million user profiles, along with 420.2 million directed relations and 105
million tweets among the users. User influence scores are obtained from
influence measurement services, Klout and PeerIndex. Our analysis reveals
interesting findings, including non-power-law influence distribution, strong
reciprocity among users in a community, the existence of homophily and
hierarchical relationships in social influences. Most importantly, we observe
that whether a user retweets a message is strongly influenced by the first of
his followees who posted that message.  To capture such an effect, we propose
the first influencer (FI) information diffusion model and show through
extensive evaluation that compared to the widely adopted independent cascade
model, the FI model is more stable and more accurate in predicting influence
spreads in Twitter communities.
\end{abstract}



\section{Introduction}
\label{sec:intro}

Recently, micro-blogging has emerged as a new medium of communication. A user can
publish short messages (or {statuses}) to spread information to his friends.
Twitter is among the most popular micro-blogging services, claiming to have
more than 500 million users by 2013~\cite{twitter500mil}.  Basic
functionalities of Twitter include disseminating {\it tweets} (short messages
with a length limit of 140 characters), updating and socializing among users. A
message can be {\it retweeted} by recipients to further spread it far beyond
the followers of its originator. Unlike other social network services that
require users to grant permissions to other users to befriend them, Twitter
employs a ``free-to-follow'' model, which allows any user to follow and get
update from others without seeking any permission. User $A$ who follows $B$ is
called $B$'s {\it follower}, while $B$ is called $A$'s {\it followee} or {\it friend}
\footnote{We use the term {\it followee} to distinguish other types of
social friendships like that of Facebook.}.

Twitter makes available application programming interfaces (APIs) that allow
open access to its data. Much work has been done towards better understanding
of the Twitter network's topological characteristics and user behaviours.
Java~\emph{et al.}~\cite{Java2007} conducted preliminary analysis on Twitter
using a small dataset of 76,000 users and 1 million tweets in 2007.  The
authors found that users cluster according to interests in topics using clique
percolation methods. Krishnamurthy~\emph{et al.}~\cite{Krishnamurthy2008}
analyzed user characteristics by the relationship between the number of
followers and followees. Zhao and Rosson~\cite{Zhao2009} qualitatively
investigated the motivation behind using Twitter. Haewoon~\emph{et
al.}~\cite{KwakWWW2010} presented the first study on the entire Twitter sphere.
Several interesting observations have been made regarding the structural
properties of the Twitter network, including non-power-law follower
distribution, short effective diameter, and low reciprocity, etc. However,
since 2011, rate limits on API calls have been enforced resulting in a drastic
reduction in the amount of research on Twitter.

{\it User influence} is defined as the ability to drive actions and provoke
interactions among others.  How to rank Twitter users based on their influence is
an active research topic.  A simple metric is the number of followers that one
has. However, recent
studies~\cite{Cha10measuringuser,WengTwitterRank2010,Bakshy2011} pointed out
that it is not a good indicator. Many researchers have striven to come up with an
intuitive and fair ranking system on Twitter. Kwak~\emph{et
al.}~\cite{KwakWWW2010}, in an effort to identify the most influential users
on Twitter, applied several ranking metrics, including the number of followers,
the number of retweets, and PageRank. It has been found that ranking results
from these metrics do not correlate well, which implies none of them is reliable.
Cha~\emph{et al.} in~\cite{Cha10measuringuser} employed another metric, i.e.,
the number of times a user is mentioned. Weng~\emph{et
al.}~\cite{WengTwitterRank2010} proposed TwitterRank, an extension of the PageRank
algorithm to measure user influence. Meanwhile, several online influence measurement
services are now available including Klout, PeerIndex, Kred, Empire Avenue, and
PROskore. These services scrape social network data, using it to create profiles
of individuals and assigning each an ``influence score''. Twitter users do not
have to register with the measurement services to have their profile evaluated,
since their information can be obtained via Twitter API.  However, if the
user registers, the service will have full access to their data and provide
more accurate measurement results. In exchange, user with high influence score
will be eligible for perks (discounted coupons, promotions, etc.) from many retailers.

In this work, we take a data-driven approach to investigate user influences in
Twitter communities identified by hashtags. The purpose of this study is
three-fold. First, we aim to understand whether characteristics previously
observed on the entire Twitter sphere (but notably of a much smaller scale than
today's Twitter network), remain valid in Twitter communities. Second, we study
the consistency between the influence scores from two major ranking services,
{\it Klout} and {\it PeerIndex}, and unravel connections between influence
scores and user relationships in Twitter communities. Third, we evaluate the
suitability of one widely adopted influence spread models, namely, the {\it
independent cascading} (IC).



From the analysis, we make several interesting observations regarding user
relationships in Twitter communities. Furthermore, we find that whether a
Twitter users retweets a message is due to the influence from the first of his
followees who posted that message. We refer to this as the {\it first
influencer} (FI) spreading model. We show through extensive evaluation that the FI model
is more accurate in prediction influence spreads in Twitter communities
compared to the classic IC model.

In summary, we make the following observations on Twitter communities:
\begin{itemize}
\item {\bf Strong reciprocity between users:} Users that share similar
interests do not randomly follow one another, but tend to follow those who follow
them forming strongly connected network components.
\item {\bf Hierarchy:} Following relationship encodes
hierarchy on Twitter, where less influential users tend to
follow those with more influence.
\item {\bf Homophily:} Homophily exists in mutual following relationships,
where users with similar influences tend to follow one another.
\item {\bf First-influencer diffusion model:} The success of information spread
in Twitter communities primarily depends on the influence of the first
information source.
\end{itemize}

The rest of this paper is organized as follows.
Section~\ref{sec:twittercrawler} describes the methodology for data collection
and the resulting datasets. Key findings from analyzing the Twitter community
datasets are presented in Section~\ref{sec:twitteranalyzing}. In
Section~\ref{sec:fimodel}, we introduce the FI model and evaluate it with two
sets of experiments.  Finally, the paper is concluded in
Section~\ref{sec:conclusion}.


\section{The Twitter Crawler}
\label{sec:twittercrawler}

Though many Twitter datasets are publicly
available~\cite{stanforddataset,www2012twitterdataset,asudataset}, they contain
little information regarding the information exchange in the network. In the present study,
we are interested not only in network structures, but also in the interactions
among users.  Most existing social datasets only contain graphs with nodes
representing users on the network and links representing follower-followee
relations. User identity is typically discarded from the dataset due to privacy
concerns.  Information exchanged among users (like tweets and messages), which is
crucial to understand and analyze influences, is considered sensitive and
cannot be published.  Therefore, we build a crawler to collect new datasets
using Twitter APIs to address the deficiency of existing datasets. 

\subsection{Implementation}
\begin{figure}[tp]
\begin{center}
\includegraphics[width=3.3in]{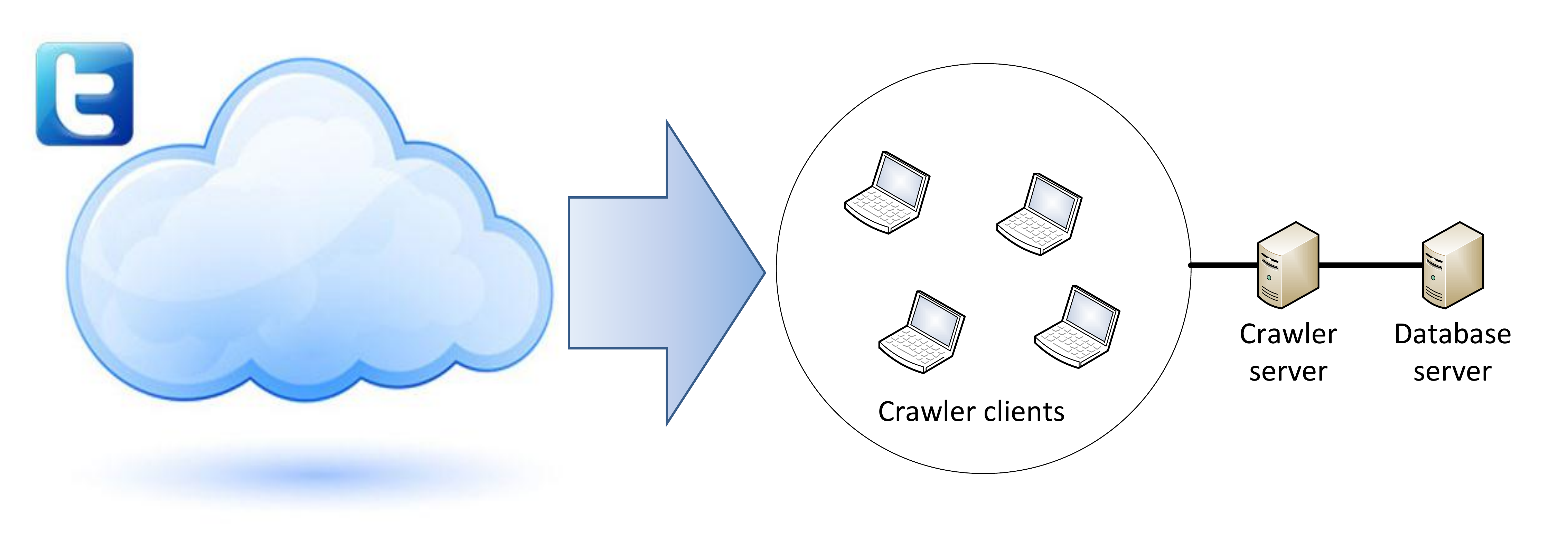}
\end{center}
\caption{The Twitter crawler}
\label{fig:twitcrawler}
\end{figure}

Twitter offers APIs to facilitate data crawling. However, due to the excessive
amount of API requests that Twitter receives, a rate-limit of 350 requests per
hour per IP address is enforced, and the whitelist
program~\cite{twitterwhitelist}  has been terminated (which allows a
whitelisted IP to make up to 20,000 requests per hour). This poses difficulty
in acquiring large amount of data since extracting the complete profile of a
user normally takes up to 3 requests.  To alleviate the above problem, we
implement a crawler in Java following the client-server model to extract both
user profiles and messages on Twitter as depicted in
Figure~\ref{fig:twitcrawler}.

Each crawler client with a different IP (either on different physical machines
or virtual machines) makes requests to crawl data from Twitter. The data is
aggregated at the crawler server. The server checks for data integrity and
correctness before storing it in the database server. We control a PC pool with
50 machines making requests continuously from October to December 2012.
\subsection{Data collection}
The initial goal of the Twitter crawler is to obtain a complete dataset
capturing nodal relations and the interactions among them.  Even with 50
machines crawling continuously, obtaining a dataset on the scale of the entire
Twitter sphere is technically infeasible.  Instead, we focus on crawling
specific communities on Twitter where users share common interests on 
trending topics.

\paragraph* {\bf Trending topics:} Twitter tracks phrases, words, and hashtags
that are often mentioned and classifies them as ``trending topics''. A hashtag
is a convention among Twitter users to create and follow a thread of discussion
by prefixing a word with a ``\#'' character. By hashtagging the word, Twitter
users create trends that may draw the community's attention. By crawling the
most popular hashtags on Twitter on many different
topics~\cite{twitter2012trends}, we obtain a diverse set of datasets that
represent the most active communities on Twitter.

\paragraph* {\bf User profiles:} Twitter profiles can be crawled from the list
of user ids that participate in each trending topic. Twitter allows public
access to a user's profile including name, location, web address, a short bio,
and the number of tweets, unless the user sets his profile to ``private''.
Persons who follow the user (followers) and those that the user follows
(followees or friends) are also listed. Note that for the sake of graph
compactness, we do not consider connections that are outside the targeted
communities. More specifically, we discard any connection and tweet of a
twitterer to and from users not in the datasets.

To this end, we obtained 20.5 million user profiles, along with 420.2 million
directed relations between followers and followees. We observe that only 8.58\%
of the users set their profile to private, preventing us from accessing their
information and relations. These users are omitted from the dataset.
Incomplete datasets as a result of the limitation on crawling rate and
processing/network resources are discarded as well.  The set of complete
datasets and their key parameters are listed in Table~\ref{tab:twitterdataset}.
In Table~\ref{tab:twitterdataset}, the network density is defined as the ratio
between the total number of edges in the network and the number of edges in a
full-connected network with the same number of vertices. Thus, a fully
connected network has a density of one; and while trees have a density of $2/n$. 

\begin{table*}[tp]
\caption{Collected datasets from Twitter.}
\small
\begin{center}
\begin{tabular}{|l||r|r|r|r|p{4cm}|}
\hline
Hashtag & Nodes & Edges & Density & Tweets/user & Trend description\\
\hline \hline
\#android & 172,817 & 1,695,021 & 1.1e-4 &  134.84 & Android phone, OS and applications\\
\hline
\#at\&t & 74,200 & 426,518 & 1.5e-4 & 67.97 & Discussions on AT\&T phone and service quality\\
\hline
\#family guy & 170,290 & 1,577,836 &  1.1e-4 & 60.25 & American animated TV show\\
\hline
\#hiphop & 93,440 & 1,862,110 & 4.2e-4 & 142.82 & Hip hop music genre\\
\hline
\#iphone & 94,928 & 501,295 & 1.1e-4 & 145.04 &  Iphone and its applications\\
\hline
\#ladygaga & 19,525 & 65,158 & 3.4e-4 & 99.64 & American female singer\\
\hline
\#marketing & 226,606 & 19,123,496 & 7.4e-4 & 215.15 &  General discussions on marketing and business\\
\hline
\#nfl & 55,200 & 703,090 & 4.6e-4 & 93.87 & American national football league\\
\hline
\#sopa & 36,993 & 474,173 & 6.9e-2 & 112.35 & U.S. bill to combat digital content piracy\\
\hline
\#teaparty & 19,772 & 3,169,181 & 1.6e-2 & 330.82 & American political party\\
\hline
\end{tabular}
\end{center}
\label{tab:twitterdataset}
\end{table*}

\paragraph* {\bf Tweets:} To collect tweets from a user, we first crawl the
tweet history. Twitter keeps the history of most recent 3,200 tweets from a
user. Older tweets are discarded. Since 3,200 tweets are insufficient to
capture active user's history, we therefore monitor each user for a one-month
period and capture all the tweets in the time period. We collect the full
text, the author, the time stamp, as well as the receiver if the tweet is a
reply. A total of 105 million tweets have been collected. %
\subsection{Removing spam tweets}

Spam tweets exist our collected datasets, most notably in the \#iphone and
\#android communities.  Removing them is therefore desirable to reduce noise
and bias in the analysis. Spam detection in Twitter is by itself an important
and active research problem.  In this work, we follow the same approach as in
~\cite{KwakWWW2010} and employ the well-known mechanism of the FireFox add-on,
Clean Tweets~\cite{cleantweets}. Clean Tweets removes tweets from users who
have been on Twitter for less than a day, and the tweets that contain three or
more trending hashtags. To this end, the average number of tweets per user from
different communities in the respective dataset is given in
Table~\ref{tab:twitterdataset}.
\subsection{Influence scores}
We crawl the influence scores of all the users in our dataset from two popular
influence measurement services:
\begin{itemize}
\item {\bf Klout~\cite{klout}:} User influence scores range from 1 to 100 with
100 being the most influential. For example, the U.S. President Barack Obama and
pop star Justin Beiber are two persons that are scored 100. Klout measures
influence mostly from Twitter data including following count, follower
count, retweets, list memberships, the influence of one's followers, etc.
\item {\bf PeerIndex~\cite{peerindex}:} It also measures one's influence
on the scale of 1 to 100. PeerIndex distinguishes itself by emphasizing its
contributions at a topic-by-topic level. The ability of users to drive
conversations and provoke interactions is reported in different topics.
\end{itemize}

\section{Twitter Community and User Influence}
\label{sec:twitteranalyzing}

In this section, we study the characteristics of the twitter communities from
the perspective of user influences.  Only results from the communities
\#ladygaga, \#sopa, \#android, and \#marketing are presented. The rest are
omitted since they are quite similar. In the subsequent analysis, we not only
unravel some Twitter-wide characteristics but also shine lights on the
differences among user communities.

\subsection{Reciprocity in following relationship}
\label{sec:reciprocity}

\begin{figure*}[t]
\begin{center}
\begin{tabular}{cccc}
\includegraphics[width=1.5in]{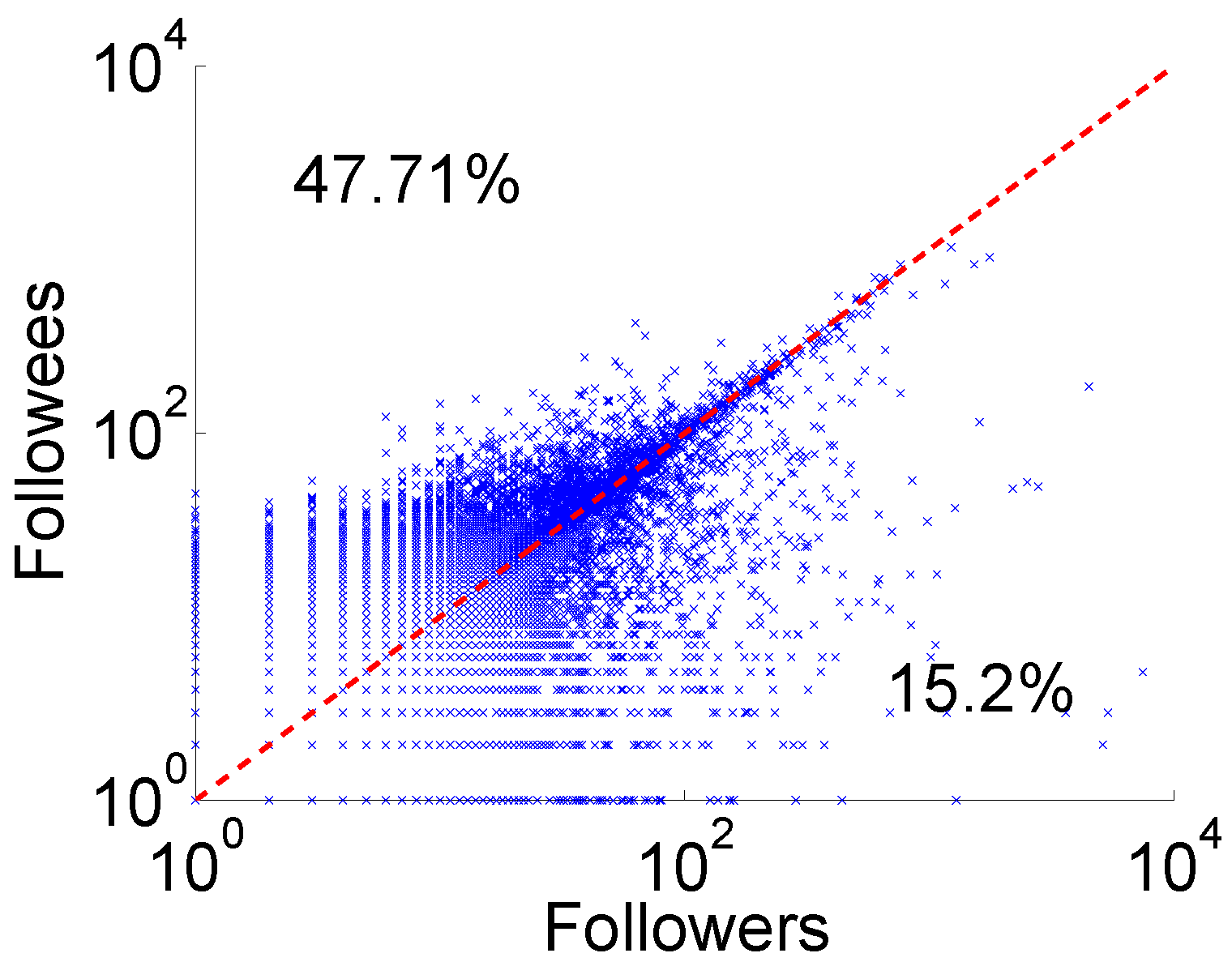} &
\includegraphics[width=1.5in]{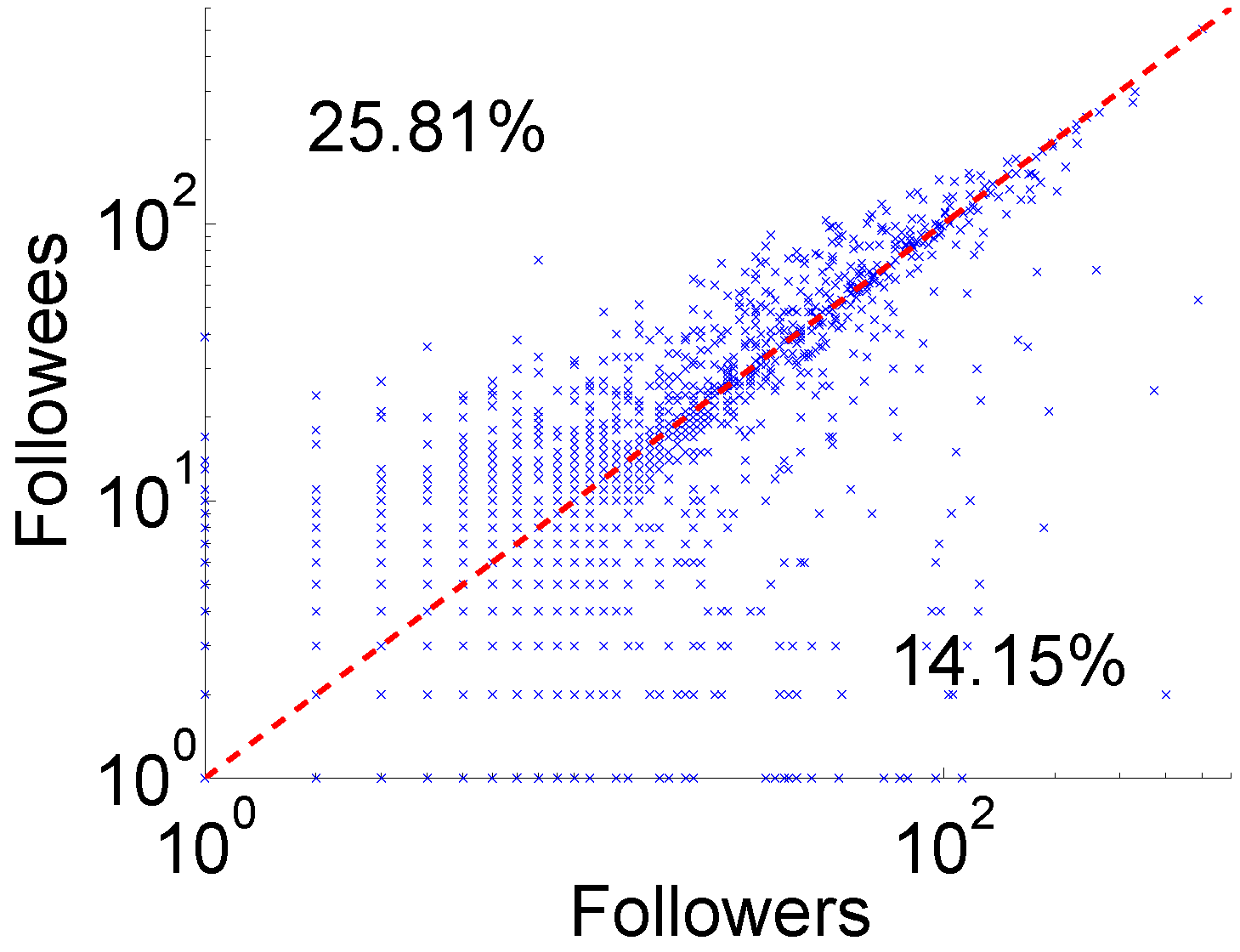} &
\includegraphics[width=1.5in]{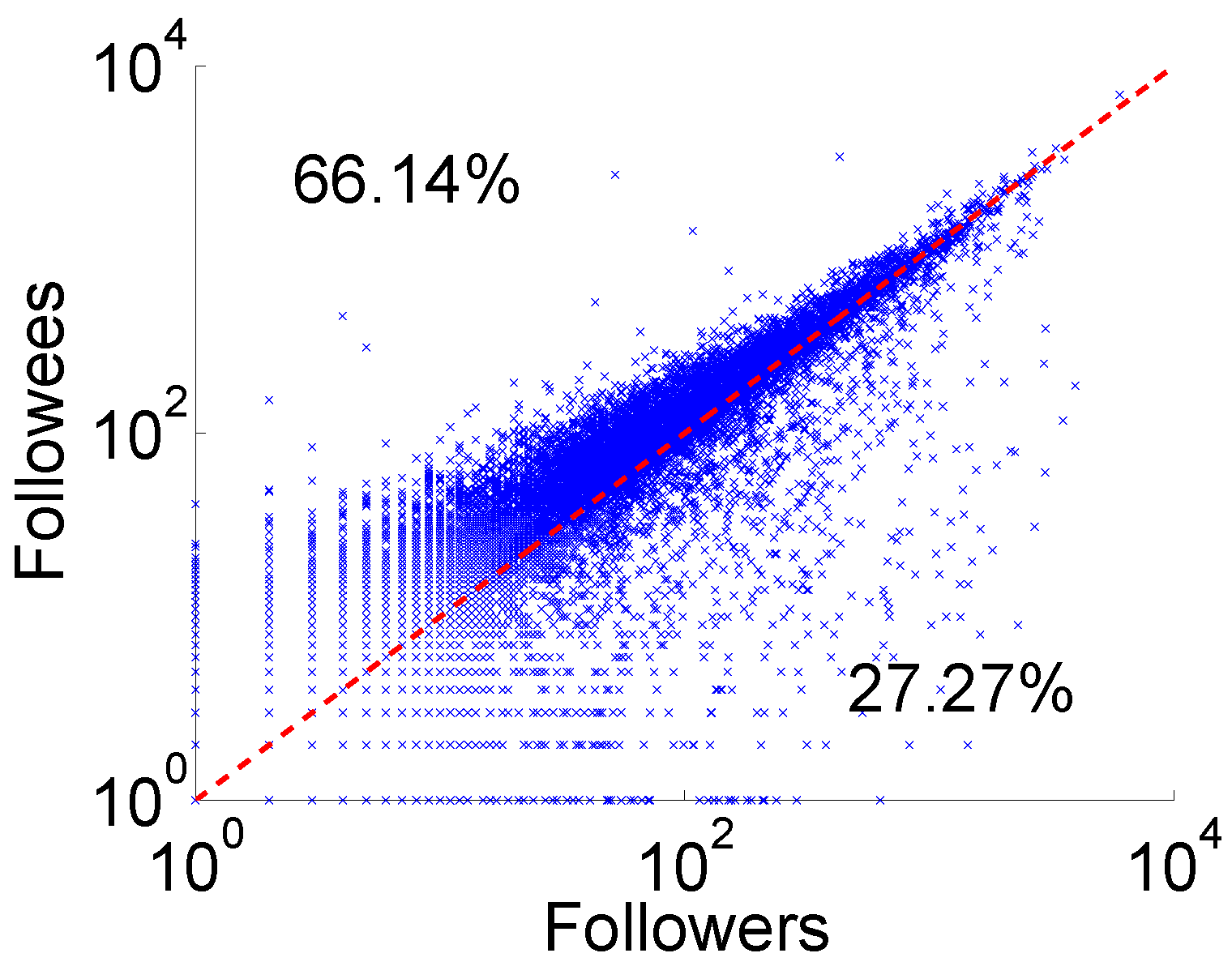} &
\includegraphics[width=1.5in]{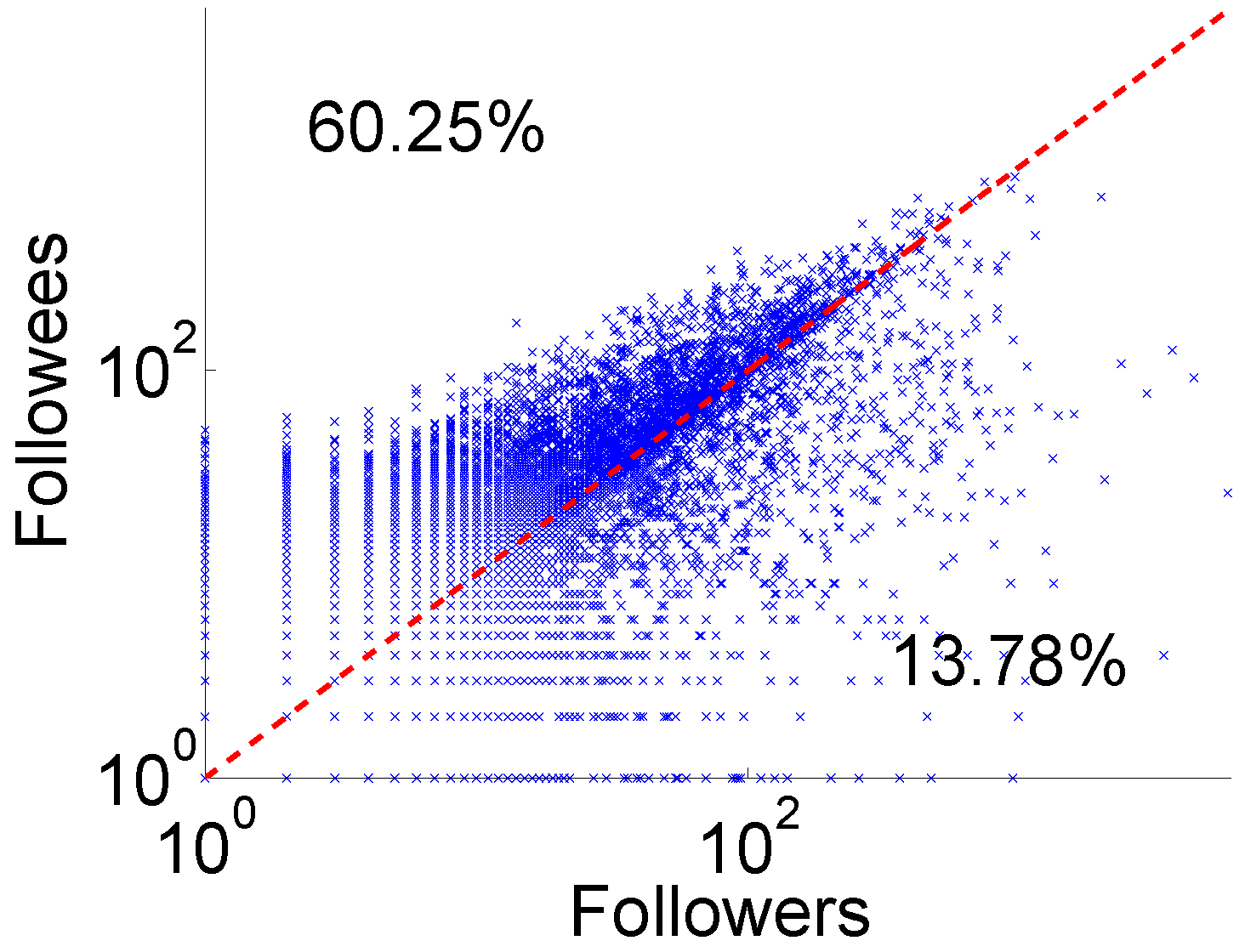}\\
{\small (a) \#android} &
{\small (b) \#ladygaga} &
{\small (c) \#marketing} &
{\small (d) \#sopa}\\
\end{tabular}
\caption{Distribution of number of followers and followees. The number above/below the diagonal line is the percentage of users with more/less followees than followers.}
\label{chap4:fig:followerfollowee}
\end{center}
\end{figure*}

We begin the analysis by presenting the basic follower/followee distribution of
different Twitter communities in Figure~\ref{chap4:fig:followerfollowee}.  The
number of followers and followees from each user is plotted in the log-log scale.
The main diagonal line (dotted line) represents the perfect reciprocity where the
number of followers is equal to the number of followees.  The number above the
diagonal line indicates the percentage of users who have more followees than
followers.  As expected, more users are above the diagonal due to the
``free-to-follow'' mechanism of Twitter.  However, we find that there are a
significant portion of twitterers with equal numbers of followers and
followees, most notably from the \#android and \#ladygaga communities (37.09\%
and 60.04\%, respectively)\footnote{Note that we only consider relationships
between users who are inside the community.}. This indicates a stronger
reciprocity in the two communities. 

\begin{figure*}[t]
\begin{center}
\begin{tabular}{cccc}
\includegraphics[width=1.5in]{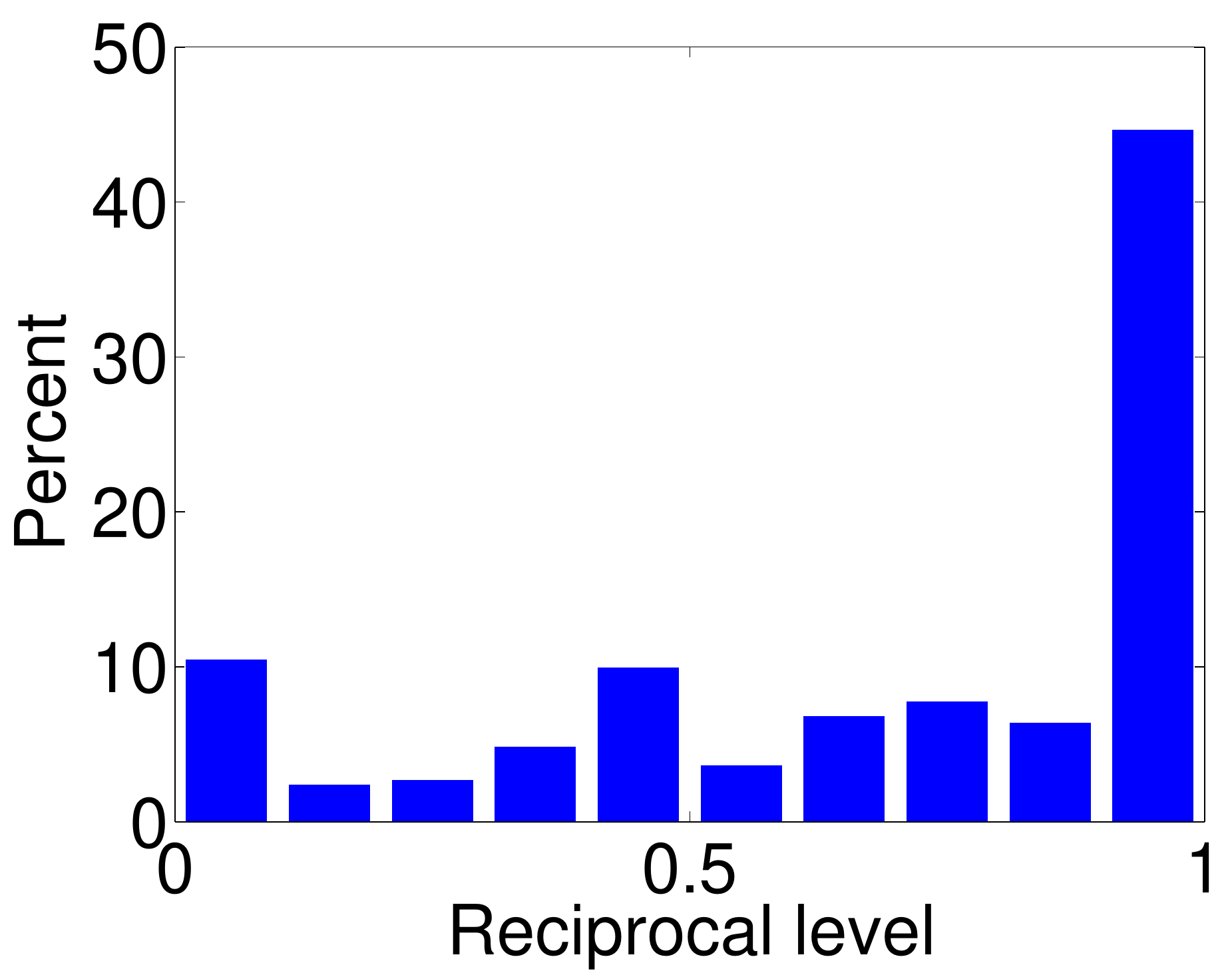} &
\includegraphics[width=1.5in]{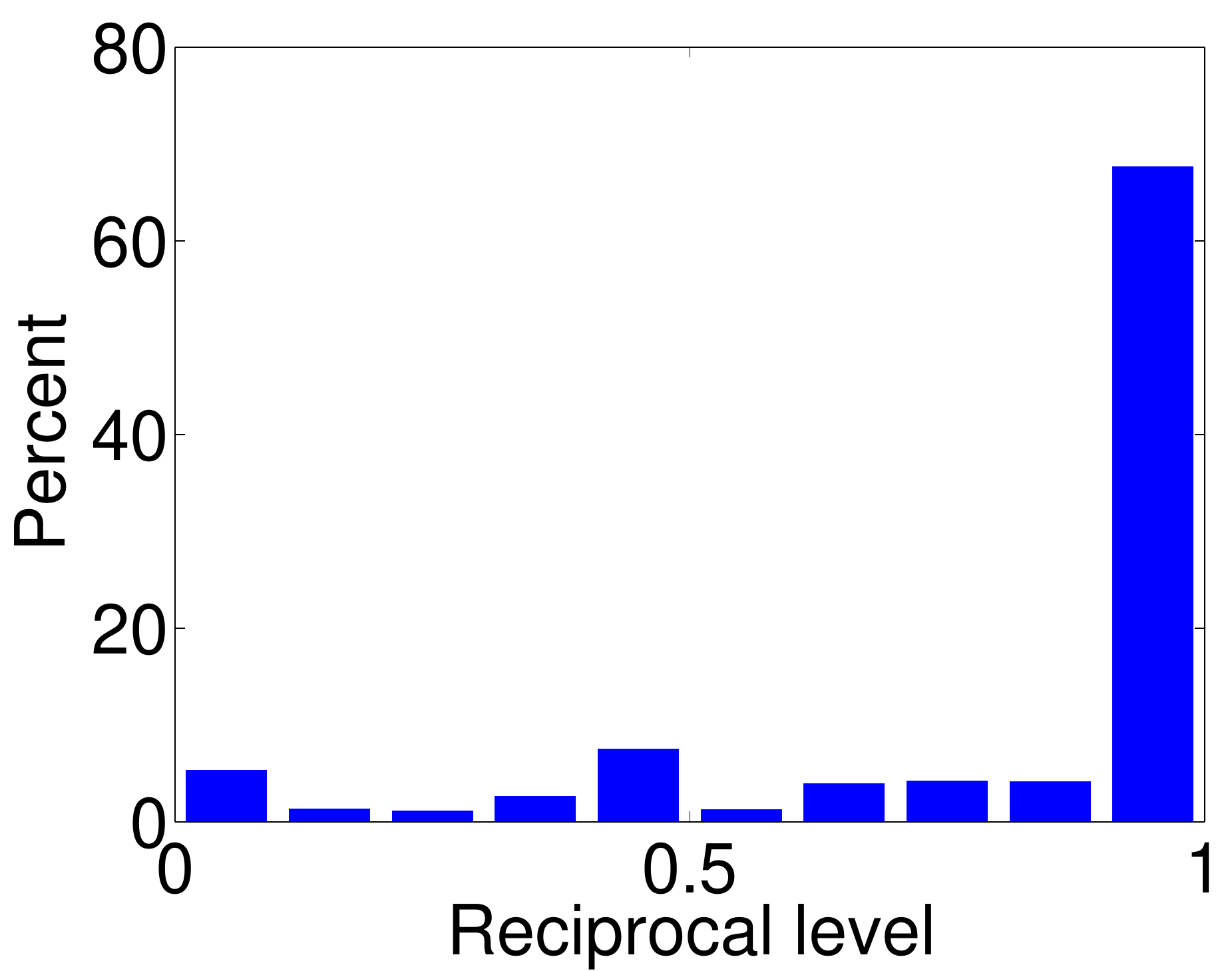} &
\includegraphics[width=1.5in]{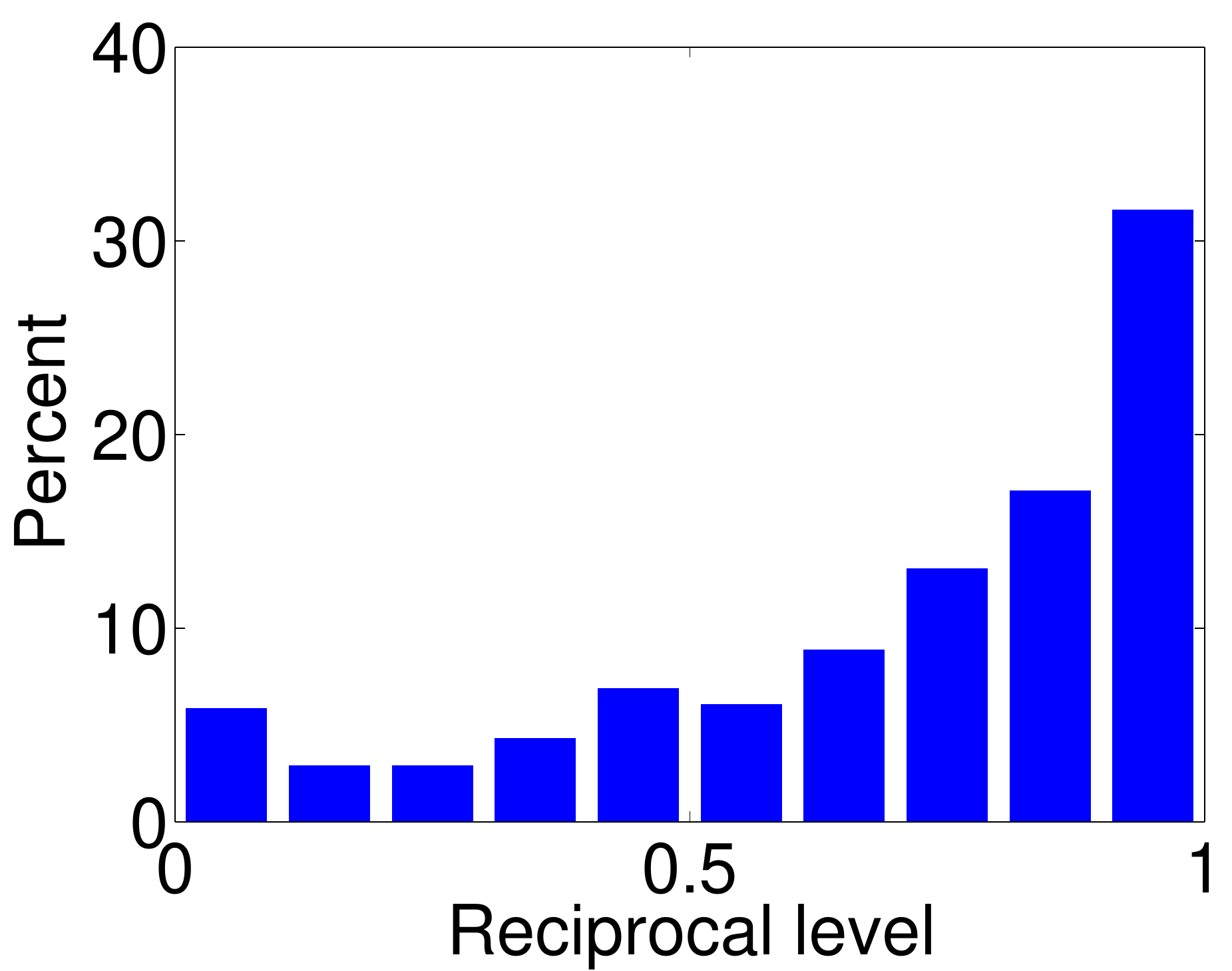}  & \includegraphics[width=1.5in]{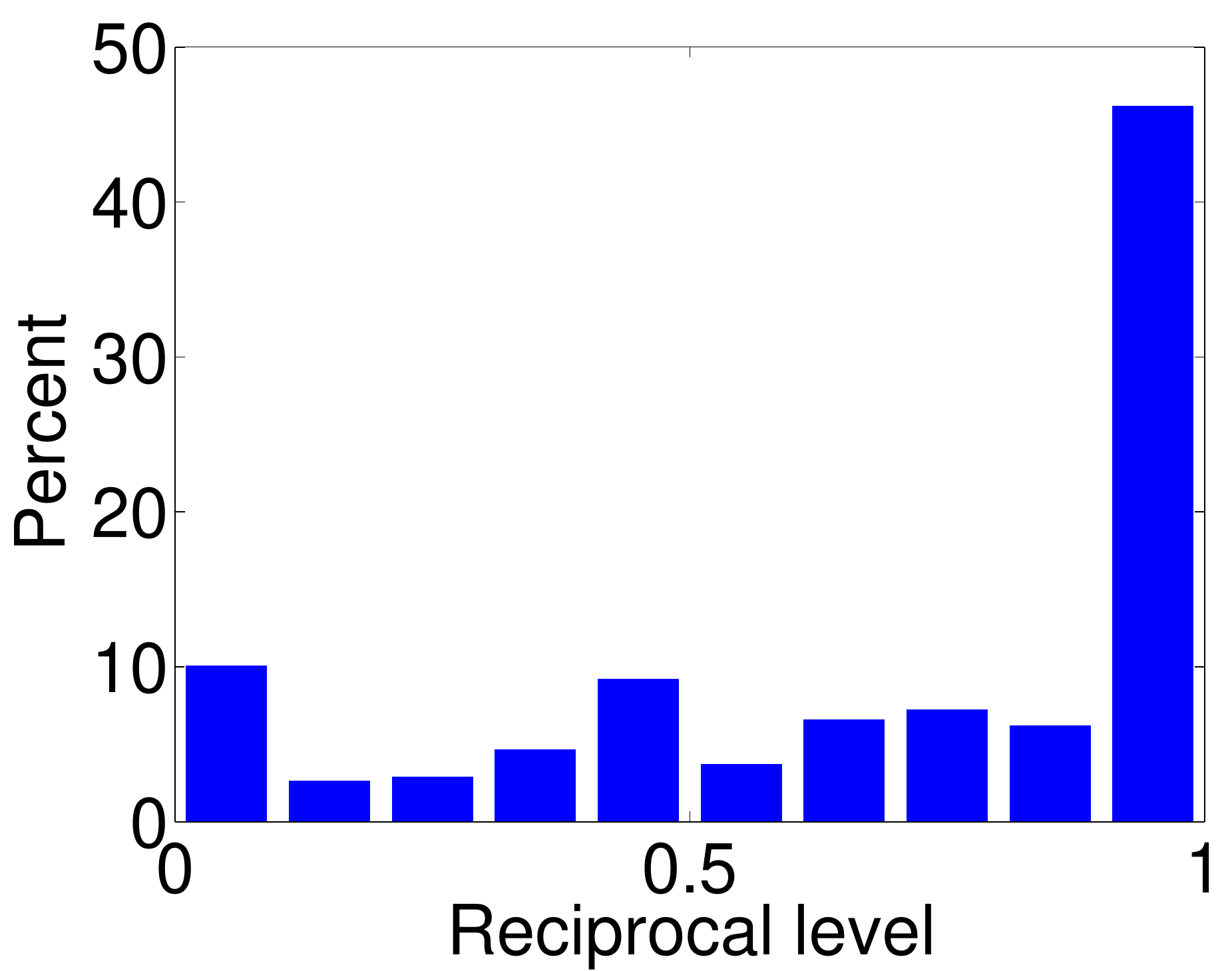}\\
{(a) \#android} &
{(b) \#ladygaga} &
{(c) \#marketing} &
{(d) \#sopa}\\
\end{tabular}
\caption{Histogram of reciprocal level.}
\label{chap4:fig:reciprocallevel}
\end{center}
\end{figure*}

Another relevant question regarding reciprocity is whether a user is likely to
follow ``back" those that follow her.  Let a {\it mutual follower} be the
follower who is also a followee. To capture mutual following,  we introduce a
new metric, {\it reciprocal level}, defined as the ratio of the number of {\it
mutual follower} to the number of followers. The histogram of {\it reciprocal
level} on four communities is shown in Figure~\ref{chap4:fig:reciprocallevel}.

From Figure~\ref{chap4:fig:reciprocallevel}, we observe that a significant portion of
users have reciprocal level 1, which means that they tend to follow who follow
them. Such a strong mutual relationship was not observed when the same study
was conducted on the scale of Twitter~\cite{KwakWWW2010}. Our results show that
at a community level, users tend to have bidirectional connections to each
other. This may be explained by the fact that users in the same community are
likely to share common interests. We also find in each community there is a
non-negligible percentage of users ($\approx 5-10\%$) with reciprocity level close to
zero. These are likely to be the community leaders who enjoy a large followings
but rarely follow back. 

\subsection{Distribution of influence scores}
\label{sec:didistribution}
\begin{figure*}[t]
\begin{center}
\begin{tabular}{cccc}
\includegraphics[width=1.5in]{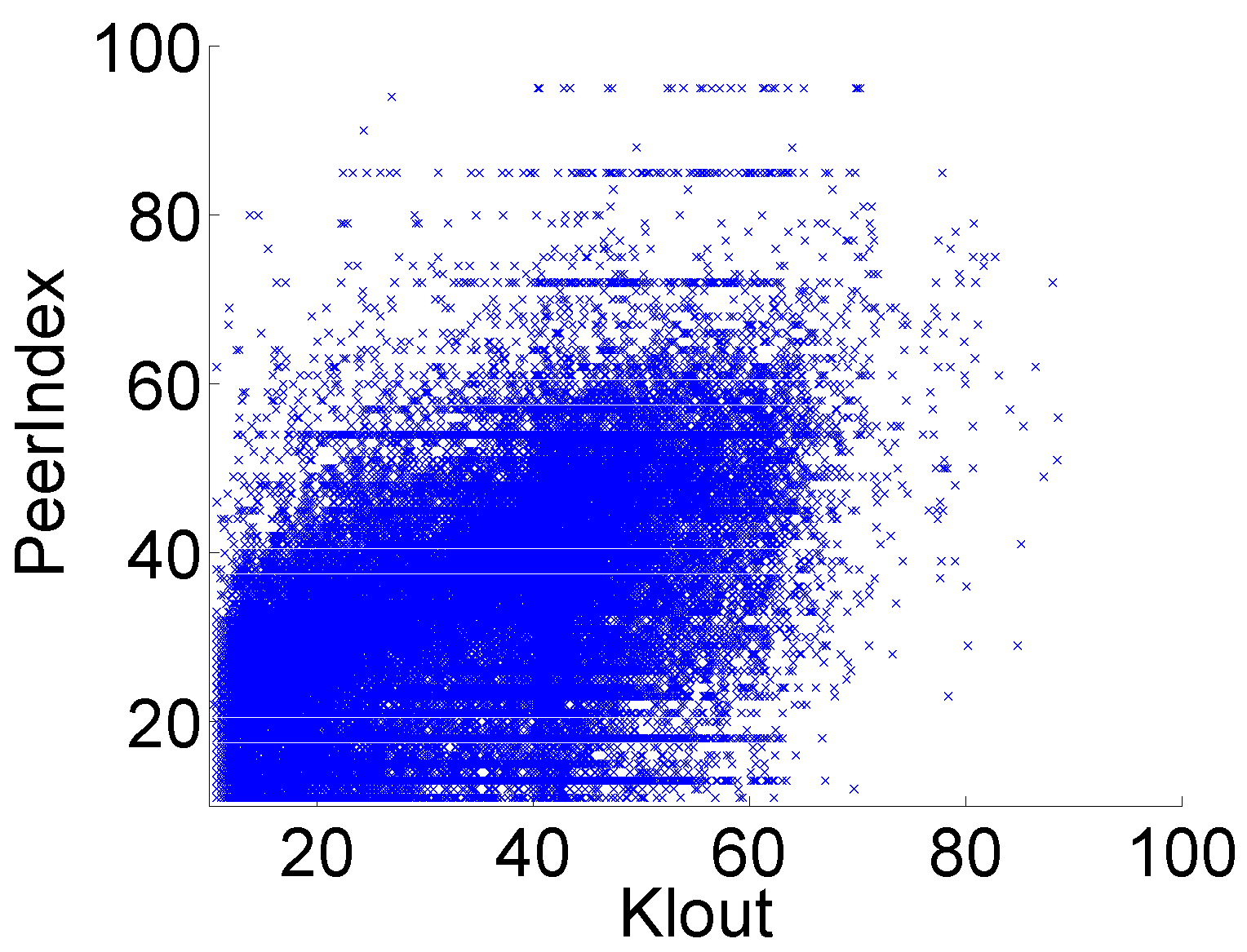} &
\includegraphics[width=1.5in]{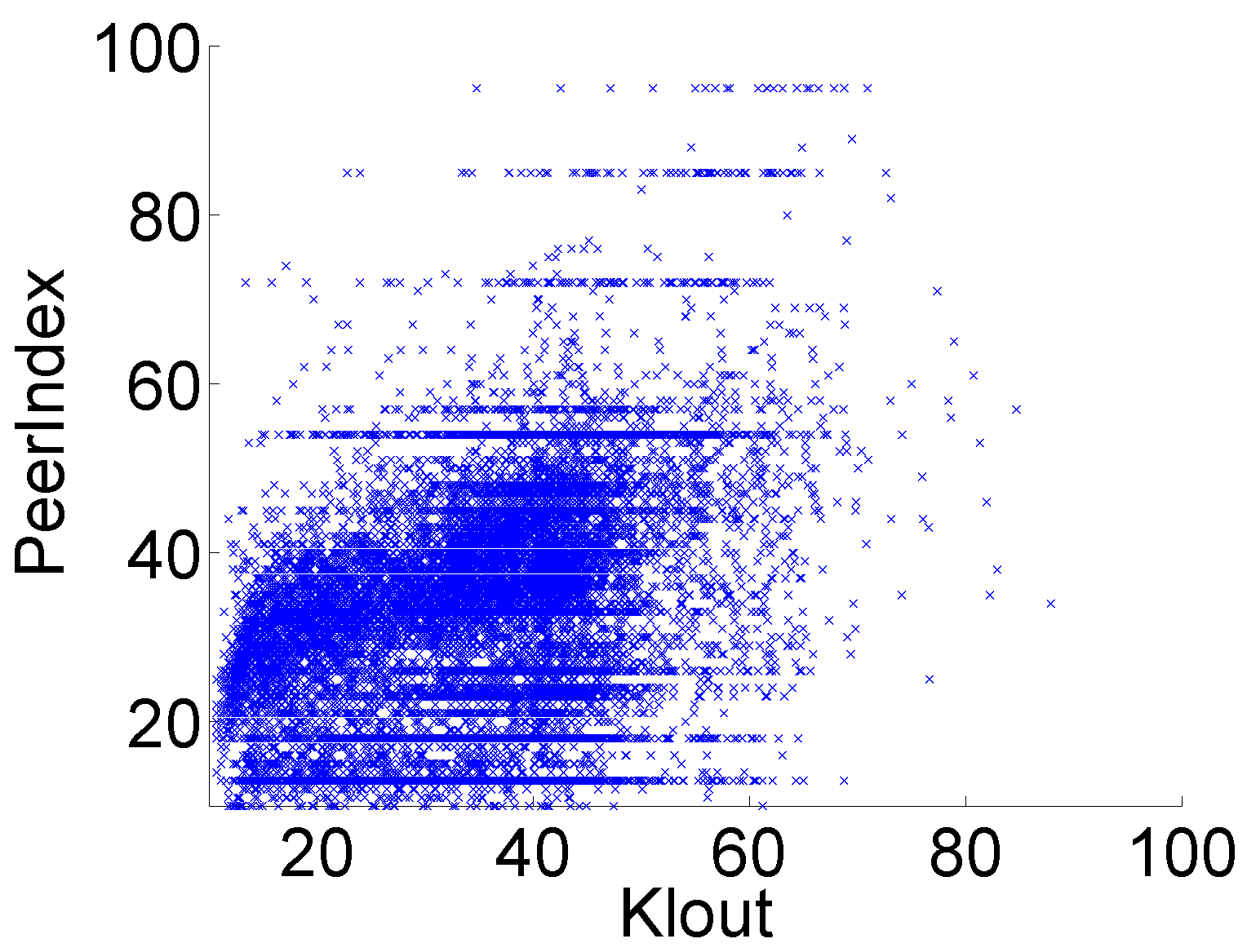} &
\includegraphics[width=1.5in]{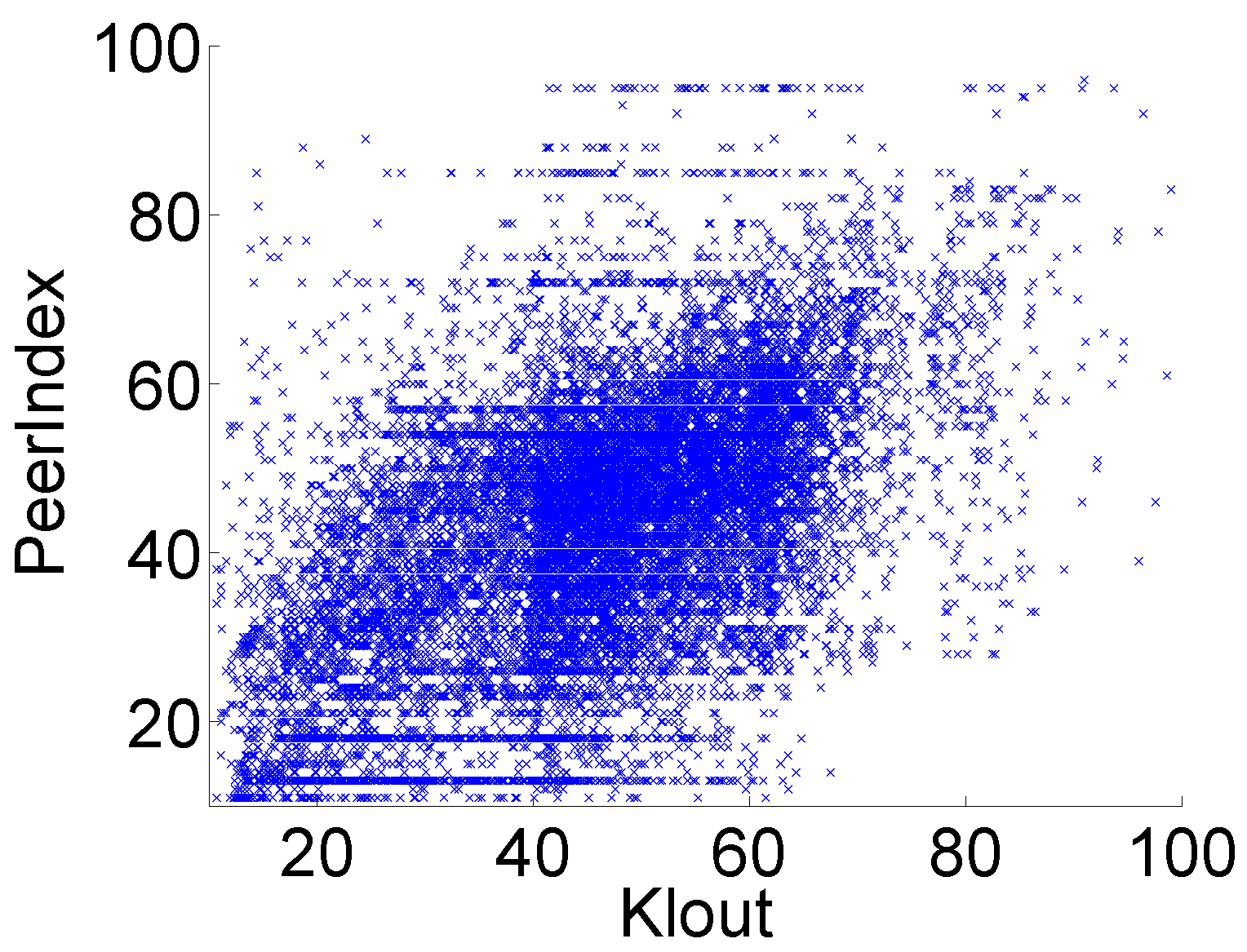} &
\includegraphics[width=1.5in]{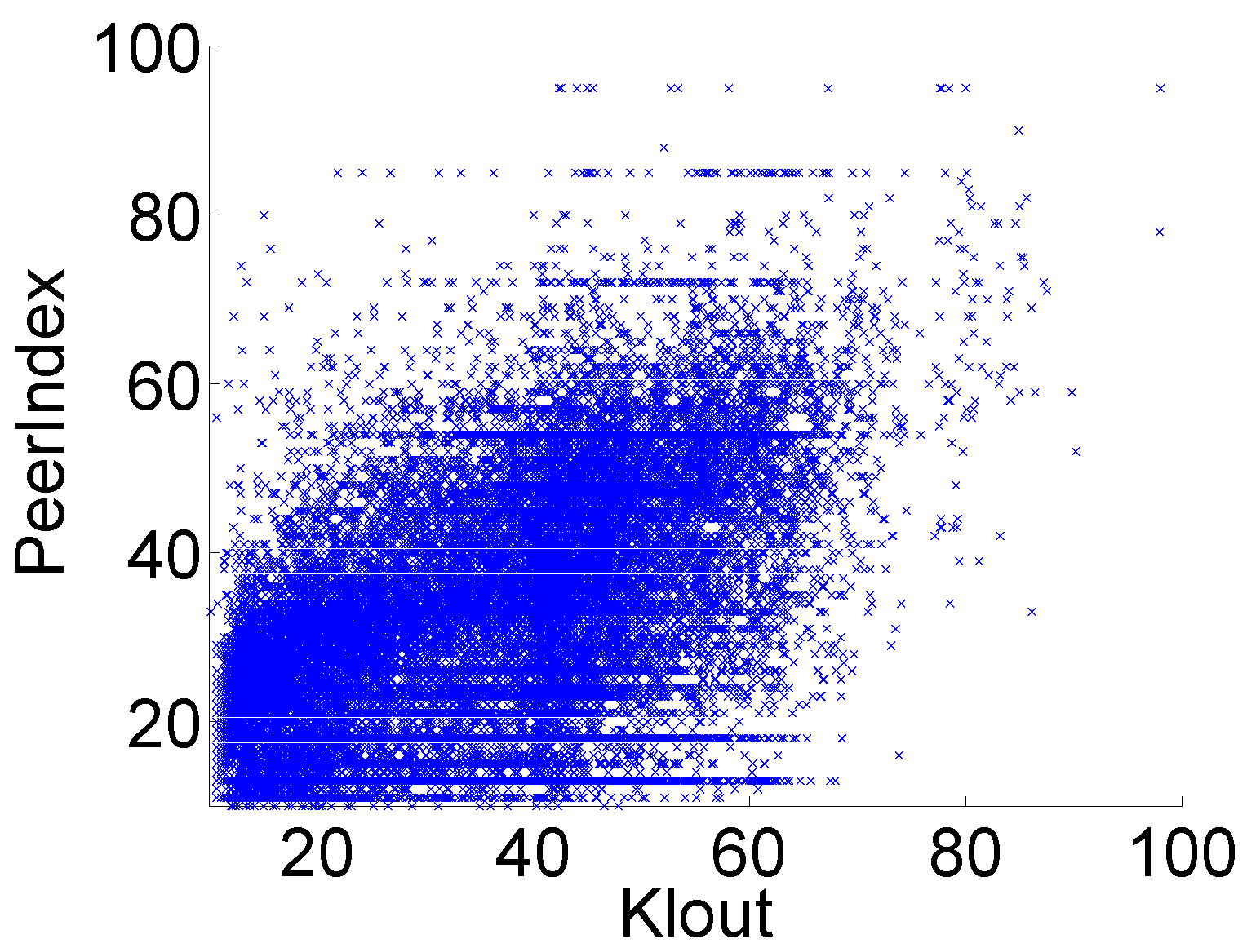}\\
{(a) \#android, $\rho = 0.5417$} &
{(b) \#ladygaga, $\rho = 0.3693$} &
{(c) \#marketing, $\rho = 0.5215$} &
{(d) \#sopa, $\rho = 0.5392$}\\
\end{tabular}
\caption{Distribution of Klout and PeerIndex scores.}
\label{chap4:fig:klpi}
\end{center}
\end{figure*}

The influence score distributions from Klout and PeerIndex are presented in
Figure~\ref{chap4:fig:klpi}. Note that a score of less than 10 is not an
indicator of weak influences, but rather, Klout and PeerIndex encounter issues
in scraping the user's data. This problem has been noted on the Klout's
developer blog~\cite{kloutblog}. As a result, in the subsequent study, we only
consider users that have both scores no less than 10 (87.2\% of all users).

From Figure~\ref{chap4:fig:klpi}, we observe that the correlation between the
scores from the two services is only moderate across all Twitter communities.
The correlation is particularly low in the \#ladygaga community with the
Pearson's correlation coefficient being $0.3693$.  Since neither of the
services publishes its ranking algorithm, we can only deduce that they use
different methods. Lack of an authoritative mechanism to measure influence, we
resort to a simplistic metric for the influence of a user by taking the average
of the two scores.  We call the new metric the {\it digital influence} (DI)
score. Note the focus of the study is not to come up with a new method in measuring
social influences rather to study the correlation between influence scores
determined by commercial ranking services with structural properties in Twitter
communities. 

\begin{figure*}[t]
\begin{center}
\begin{tabular}{cccc}
\includegraphics[width=1.5in]{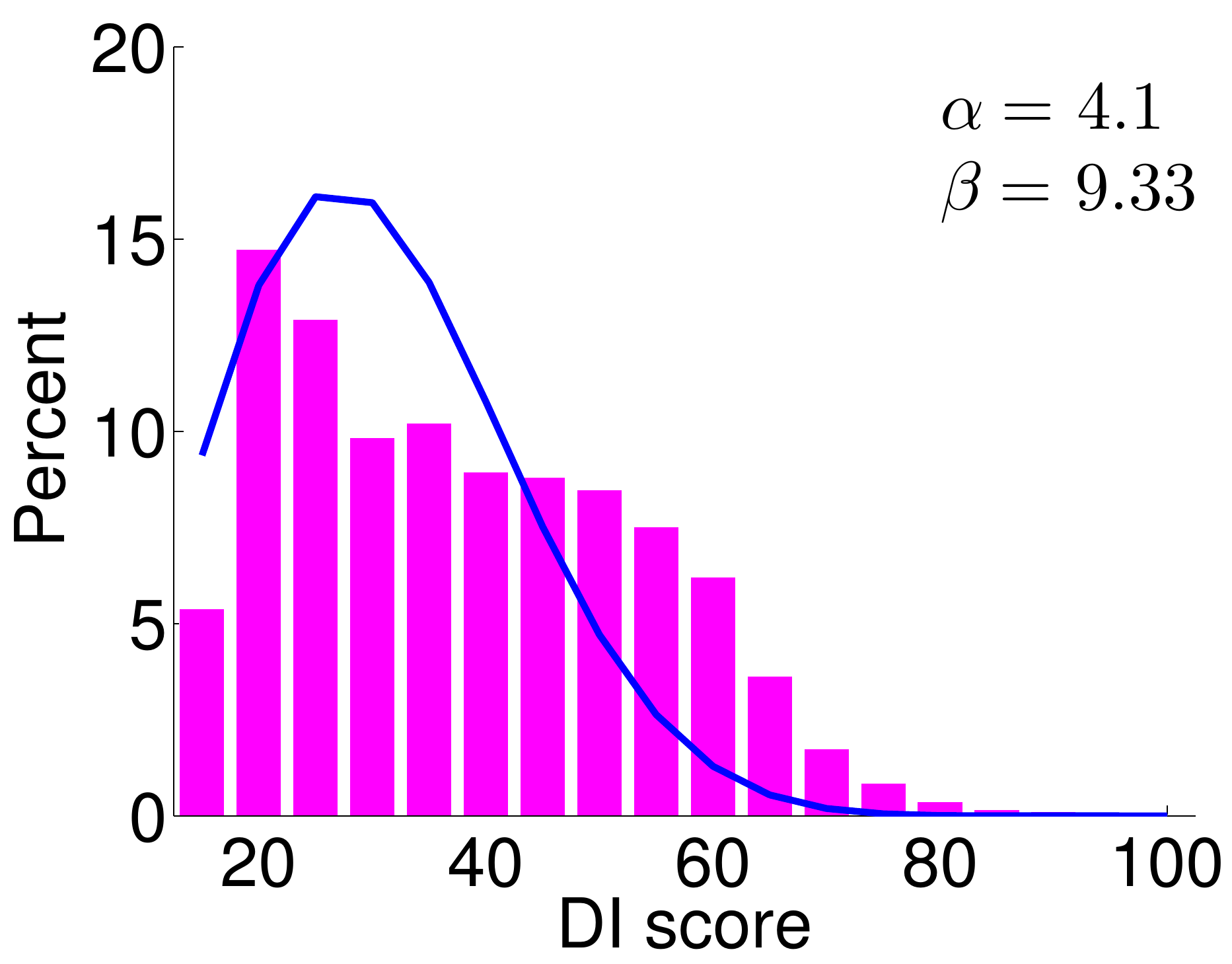} &
\includegraphics[width=1.5in]{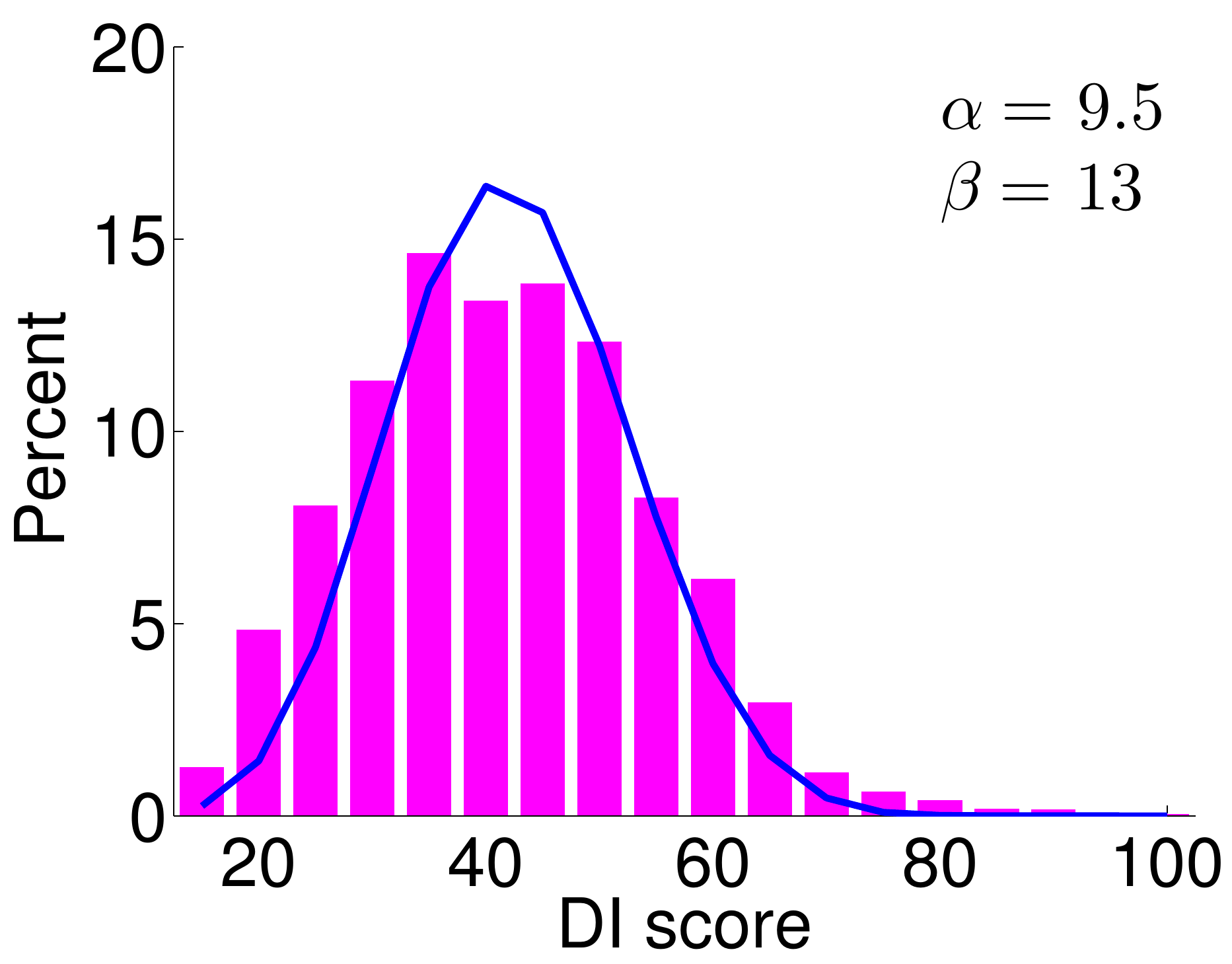} &
\includegraphics[width=1.5in]{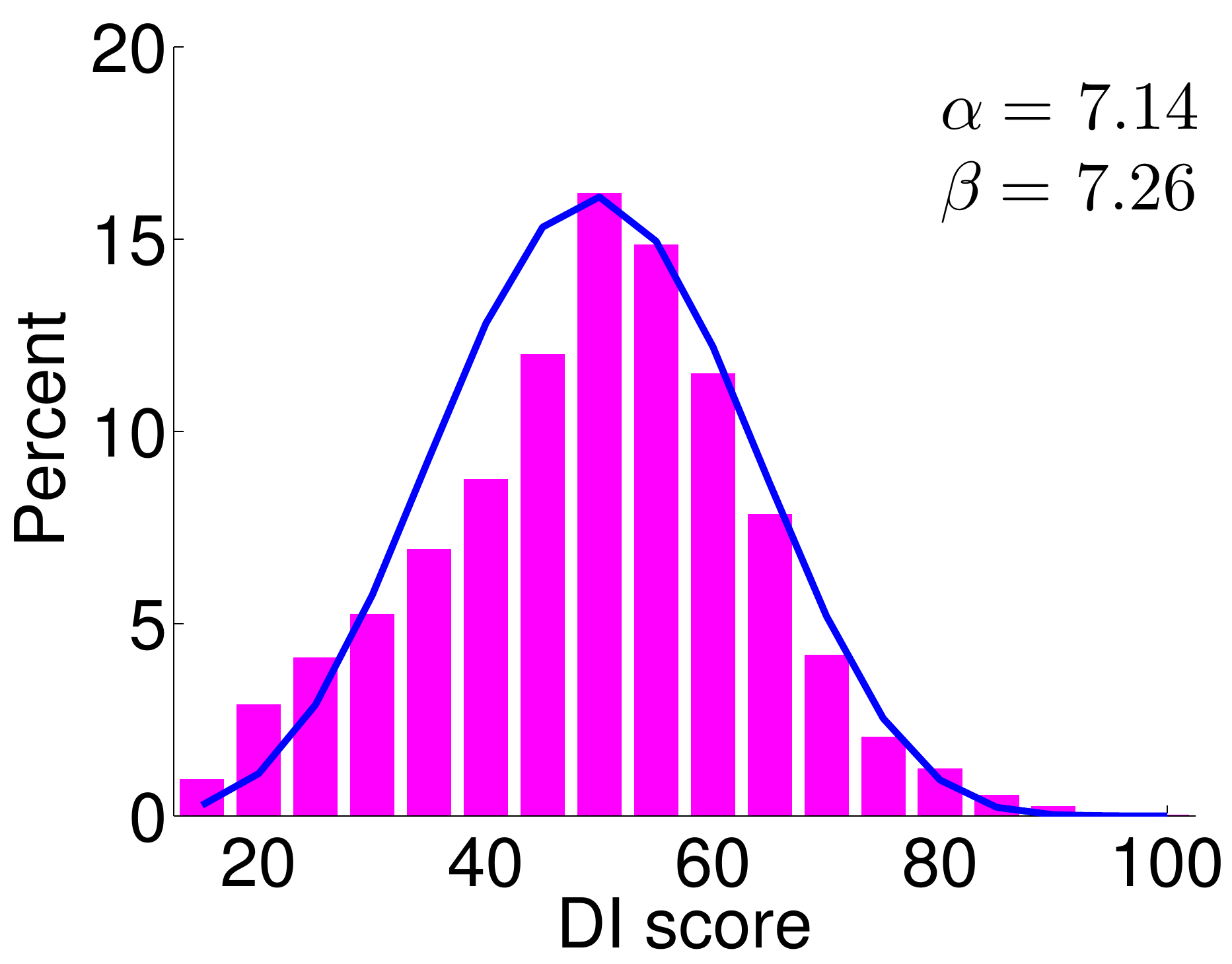} &
\includegraphics[width=1.5in]{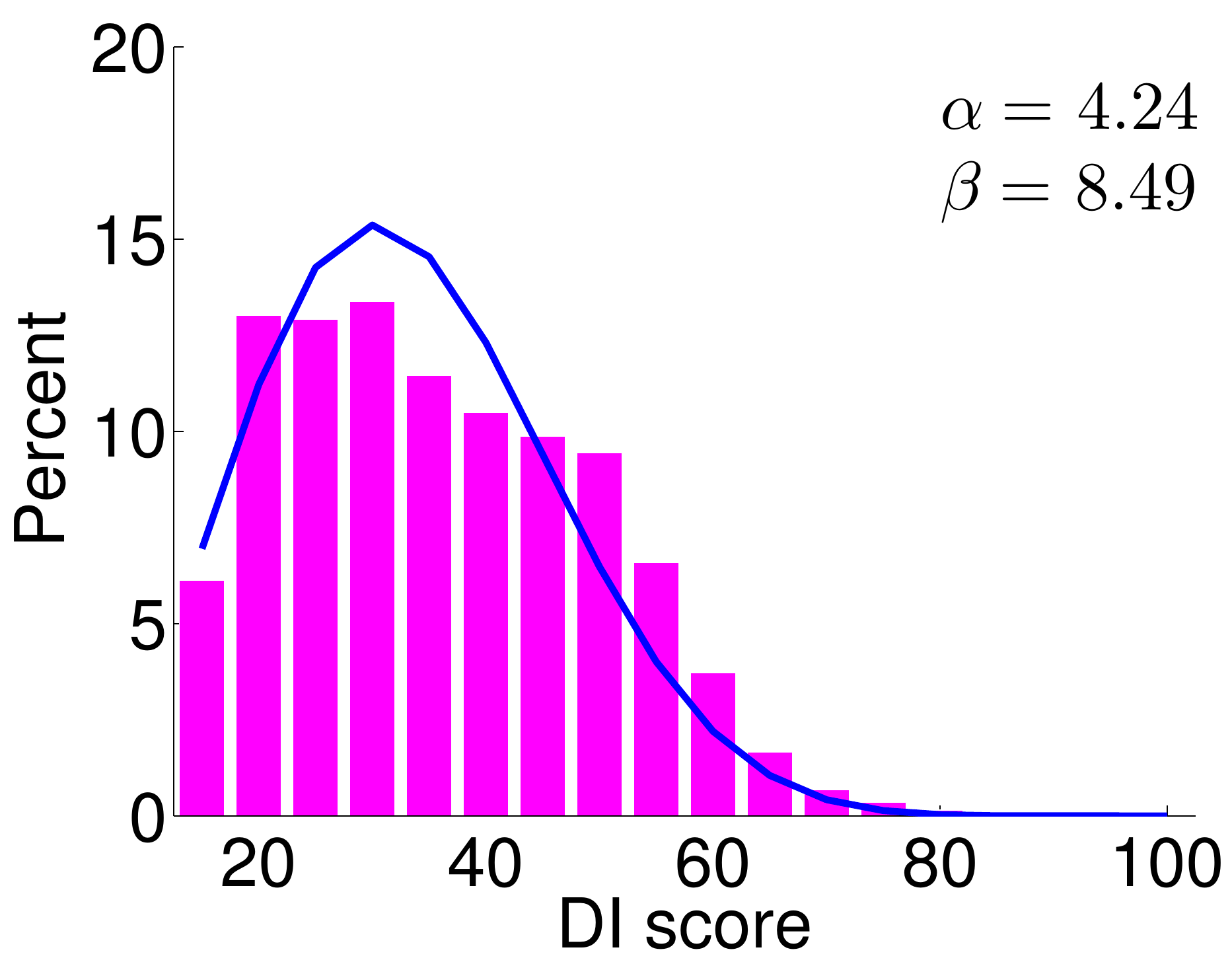}\\
{(a) \#android} &
{(b) \#ladygaga} &
{(c) \#marketing} &
{(d) \#sopa}\\
\end{tabular}
\caption{Histogram of DI score and fitting using beta distribution.}
\label{chap4:fig:di}
\end{center}
\end{figure*}

The histogram of the DI score and the maximum likelihood fitting of a beta
distribution are depicted in Figure~\ref{chap4:fig:di}.  We find that the DI
score does not follow a power law distribution, but roughly a beta
distribution with two shape parameters $2 < \alpha < \beta$ and the mean value
around 30 to 40. We also observe that the mean DI score varies from one
community to another as summarized in Table~\ref{chap4:table:meandi}. The mean
DI score is higher in the ``\#marketing'' community since it includes many
business people and mass media entities, who tend to have strong influence on
others. In contrast, the average DI score is lower in the ``\#android''
community. Our dataset reveals that most tweets that contain the ``\#android''
tag are from people who play Android games. They often post tweets containing
information of the game with the ``\#android'' tag to receive perks or bonuses
from game providers.  These tweets would mostly be discarded by other
twitterers resulting in a low average DI score in the ``\#android'' community.

\begin{table}[h]
\caption{Mean DI score in different communities.}
\small
\begin{center}
\begin{tabular}{|c||c|c|c|c|}
\hline
Dataset & \#android & \#ladygaga & \#marketing & \#sopa\\
\hline
Mean DI & 29.15 & 34.33 & 45.04 & 33.25\\
\hline
\end{tabular}
\end{center}
\label{chap4:table:meandi}
\end{table}

\subsection{Hierarchy}
\label{sec:hierarchy}

Social hierarchy or stratification among humans is a well studied concept in
sociology. Online social networks with their tremendous amount of available
data give rise to new opportunities to study the social hierarchy for networks
of different types and scales.

Although there is no formal definition of stratification, recent studies show
that hierarchy does exist in many online social networks, including Twitter.
Researchers in~\cite{Rowe2007,Maiya2009,Gupte2011} hypothesize that people form
connections in a social network based on their perceived social hierarchy. For
instance, $A$ following $B$ is a reflection that $B$'s social rank is likely higher
than $A$.

In absence of the ground truth of social hierarchy, we make the simplified
assumption that a person's influence score is positively correlated with his
rank. In other words, a person that ranks higher in the social hierarchy tends
to have a higher influence score, and vice versa. Furthermore, a user of a high
social rank is unlikely to follow those with lower social ranks.  To verify the
later hypothesis, we analyze the DI scores of a user's followers and followees.

Let $N_{out}(u)$ and $N_{in}(u)$ be the set of node $u$'s followers and
followees, respectively. We define $\Delta_r(u)$ and $\Delta_e(u)$ to be the
difference between the DI score of $u$ and the mean DI score of $u$'s followers
and followees, respectively.

\begin{center}
$\Delta_r(u) = DI(u) - \frac{\sum_{\forall v \in N_{out}(u)}DI(v)}{|N_{out}(u)|}$ and $\Delta_e(u) = DI(u) - \frac{\sum_{\forall v \in N_{in}(u)}DI(v)}{|N_{in}(u)|}.$
\end{center}

\begin{figure*}[t]
\begin{center}
\begin{tabular}{cccc}
\includegraphics[width=1.5in]{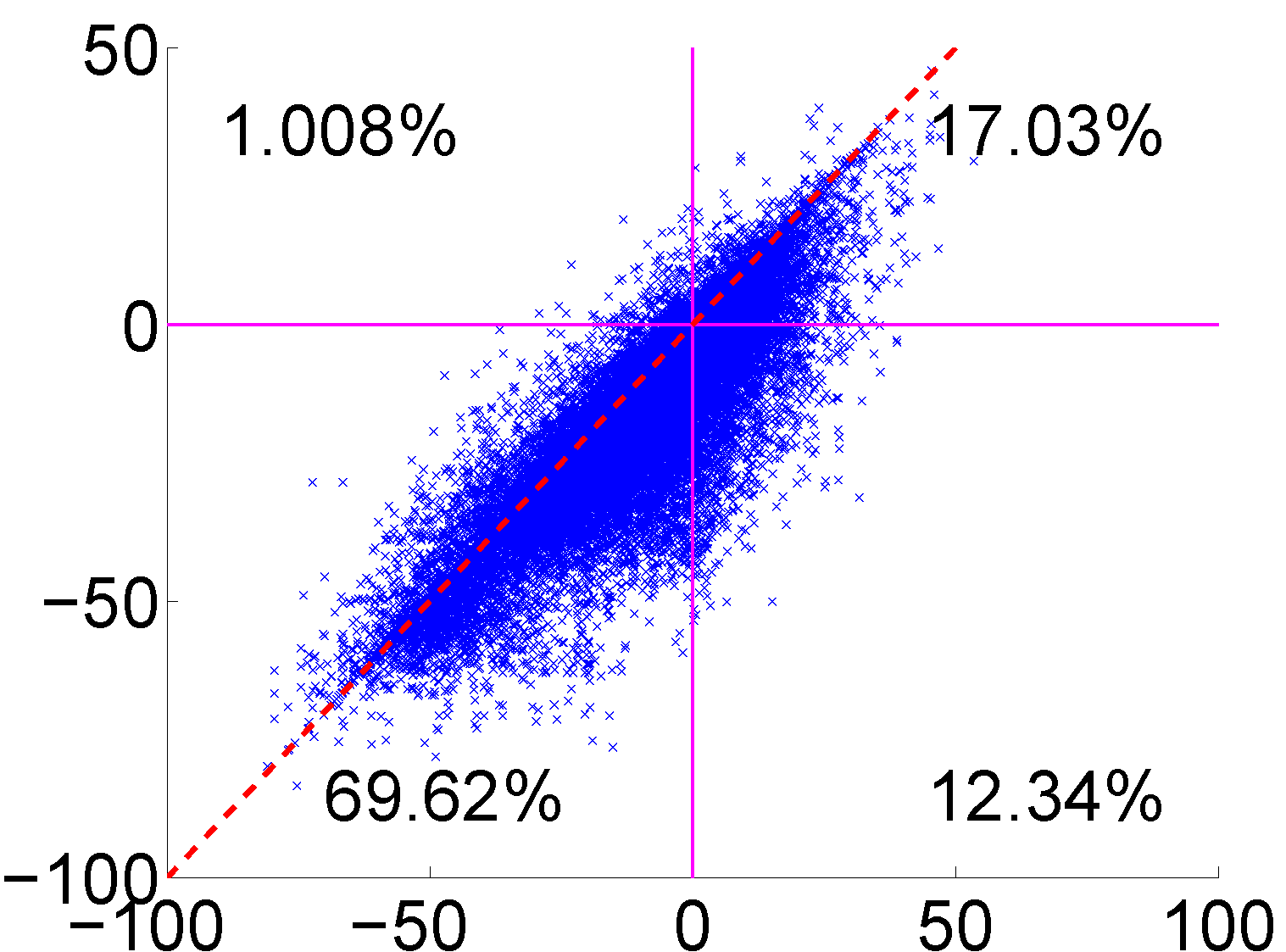} &
\includegraphics[width=1.5in]{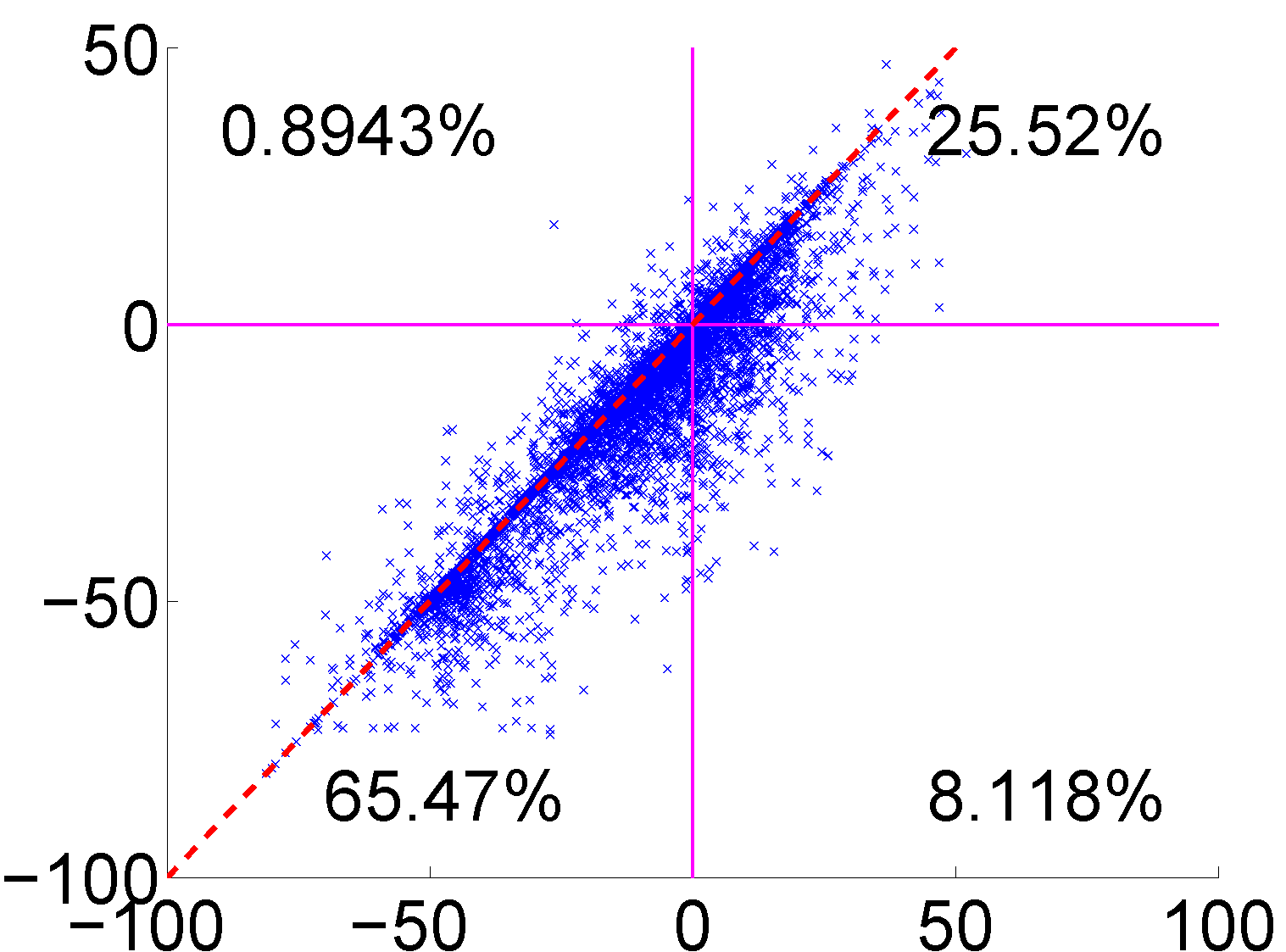} &
\includegraphics[width=1.5in]{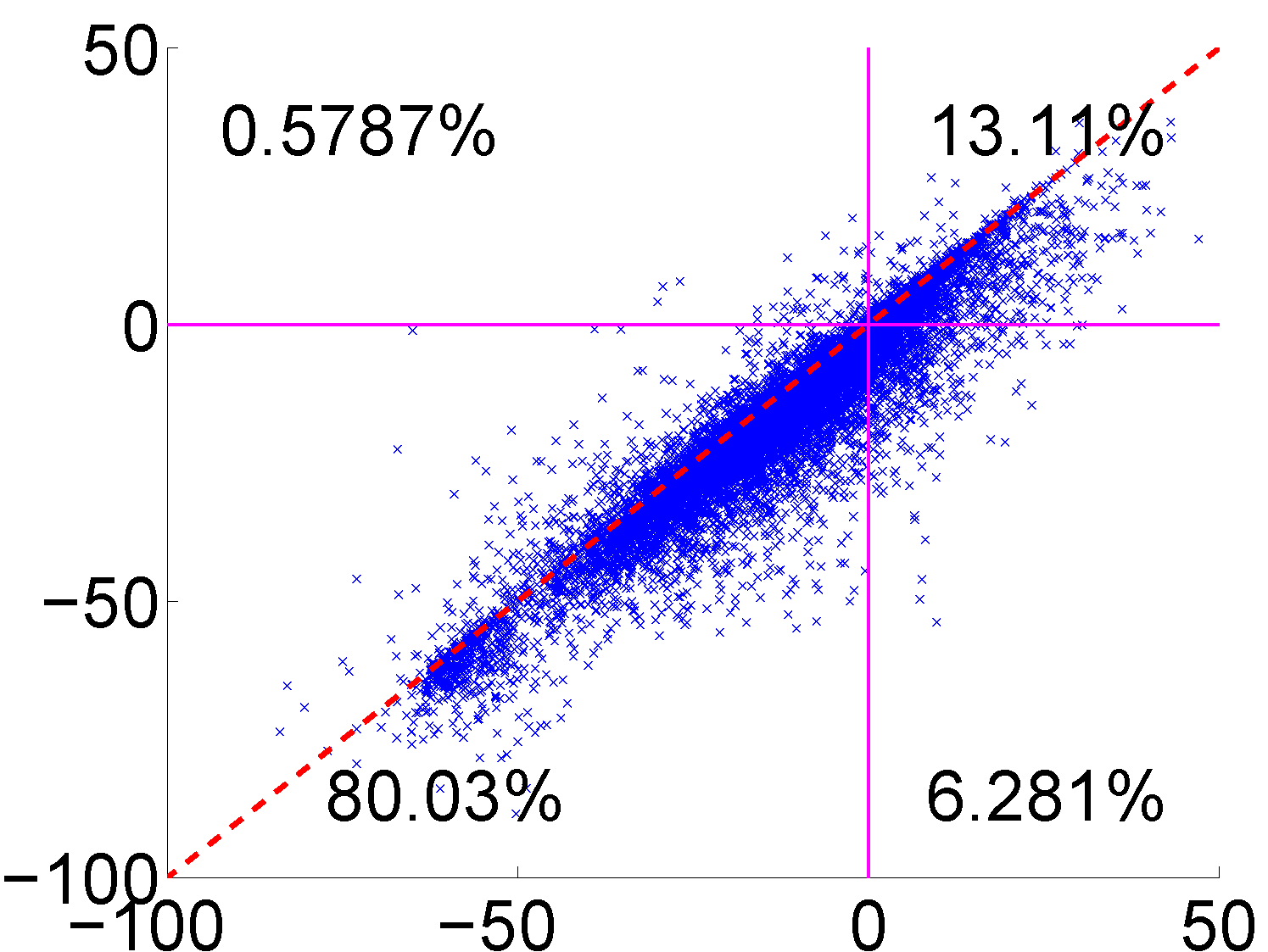} &
\includegraphics[width=1.5in]{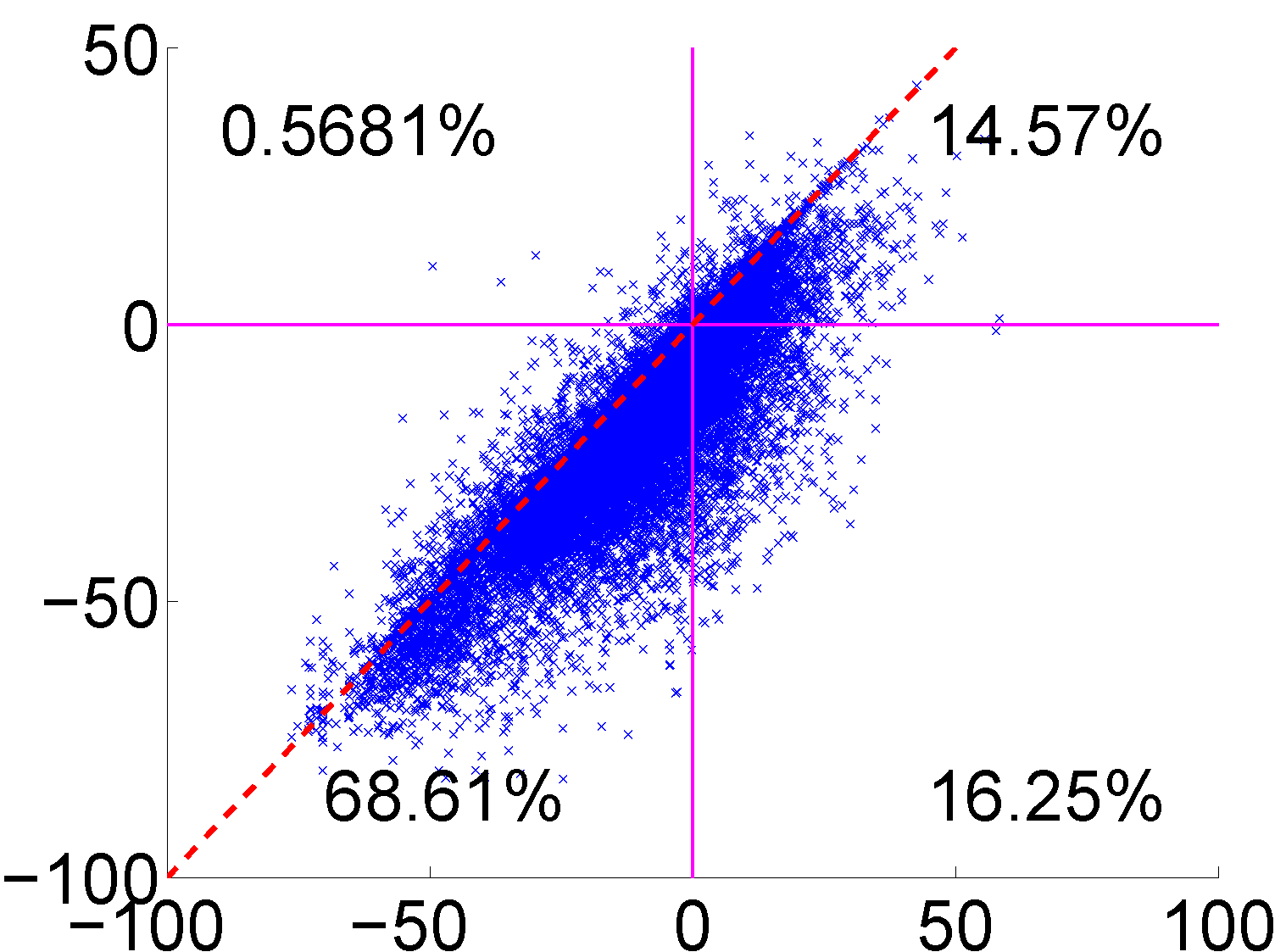}\\
{(a) \#android} &
{(b) \#ladygaga} &
{(c) \#marketing} &
{(d) \#sopa}\\
\end{tabular}
\caption{Distribution of $\Delta_r$ and $\Delta_e$. $x$-axis indicates $\Delta_r$ and $y$-axis indicates $\Delta_e$.}
\label{chap4:fig:hierarchy}
\end{center}
\end{figure*}

We calculate $\Delta_r(u)$ and $\Delta_e(u)$ for all users in our datasets.
Those that do not have any follower or followee are discarded.  Results
from Figure~\ref{chap4:fig:hierarchy} show that the majority of users have
$\Delta_r > \Delta_e$, which means the average score of their followees is higher
than that of their followers. This indicates that the following relationship in
Twitter encapsulates hierarchical information, and a user's followees tend to be
more influential than her followers.

Before conducting the study, we expect that most users should have $\Delta_r >
0 > \Delta_e$. In other words, a user is at lower social rank than its
followees. However, in our datasets, majority of users have $0 > \Delta_r >
\Delta_e$. This could be attributed to two factors.  First, DI score follows
the beta distribution with shaping parameters $\beta > \alpha$ as previously
illustrated in Figure~\ref{chap4:fig:di}.  Therefore, more users have DI score
less than the mean score of their communities. Second, removal of users who do
not have any follower or followee can also lead to skews in the distributions
of $\Delta_r$ and $\Delta_e$.



\subsection{Homophily}
\label{sec:homophily}

{\it Homophily} is the phenomenon where people's social networks ``are
homogeneous with regard to many sociodemographic, behavioural, and interpersonal
characteristics''~\cite{birdofafeather}. In the context of Twitter, homophily
implies that there are stronger connections between those who are ``socially
equal''. Understanding homophily can help us build better user models for
personalization and recommendation systems. Many previous
studies~\cite{WengTwitterRank2010} have verified homophily in Twitter along
many dimensions, such as age, location, occupation, topical interest, and
expertise, etc. In this section, we study homophily from the perspective of
user influence.

We test the hypothesis that users with similar influence are likely to be
mutual followers. Let $N_{re}(u)$ be the set of reciprocal followers of $u$
(those that are both $u$'s follower and followee) and $N_{nre}(u)$ be the rest
of $u$'s followers who are not in $N_{re}(u)$. We have $N_{re}(u) = \{v|v\in
N_{out}(u) \wedge u\in N_{out}\}$ and $N_{nre}(u) = \{v|v\in N_{out}(u) \wedge
v \not\in N_{re}(u)\}$. Define $\Delta_{re}(u)$ and $\Delta_{nre}(u)$ as the
average score distance of $u$ to users in $N_{re}(u)$ and $N_{nre}(u)$,
respectively:

\begin{center}
$\Delta_{re}(u) = \frac{\sum_{\forall v \in N_{re}(u)}{|DI(u) - DI(v)|}}{|N_{re}(u)|}$ and $\Delta_{nre}(u) = \frac{\sum_{\forall v \in N_{nre}(u)}{|DI(u) - DI(v)|}}{|N_{nre}(u)|}.$
\end{center}

\begin{figure*}[t]
\begin{center}
\begin{tabular}{cccc}
\includegraphics[width=1.5in]{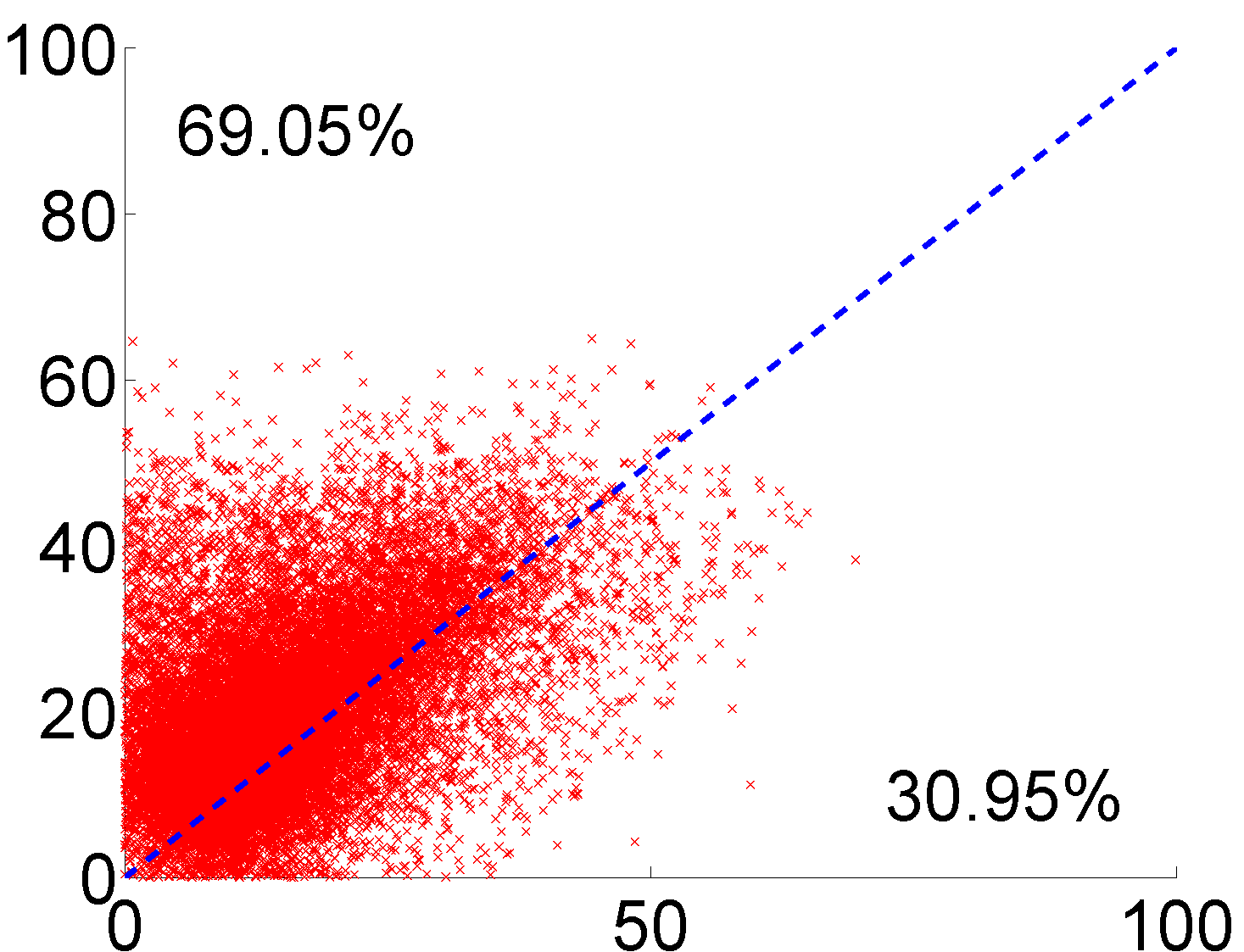} &
\includegraphics[width=1.5in]{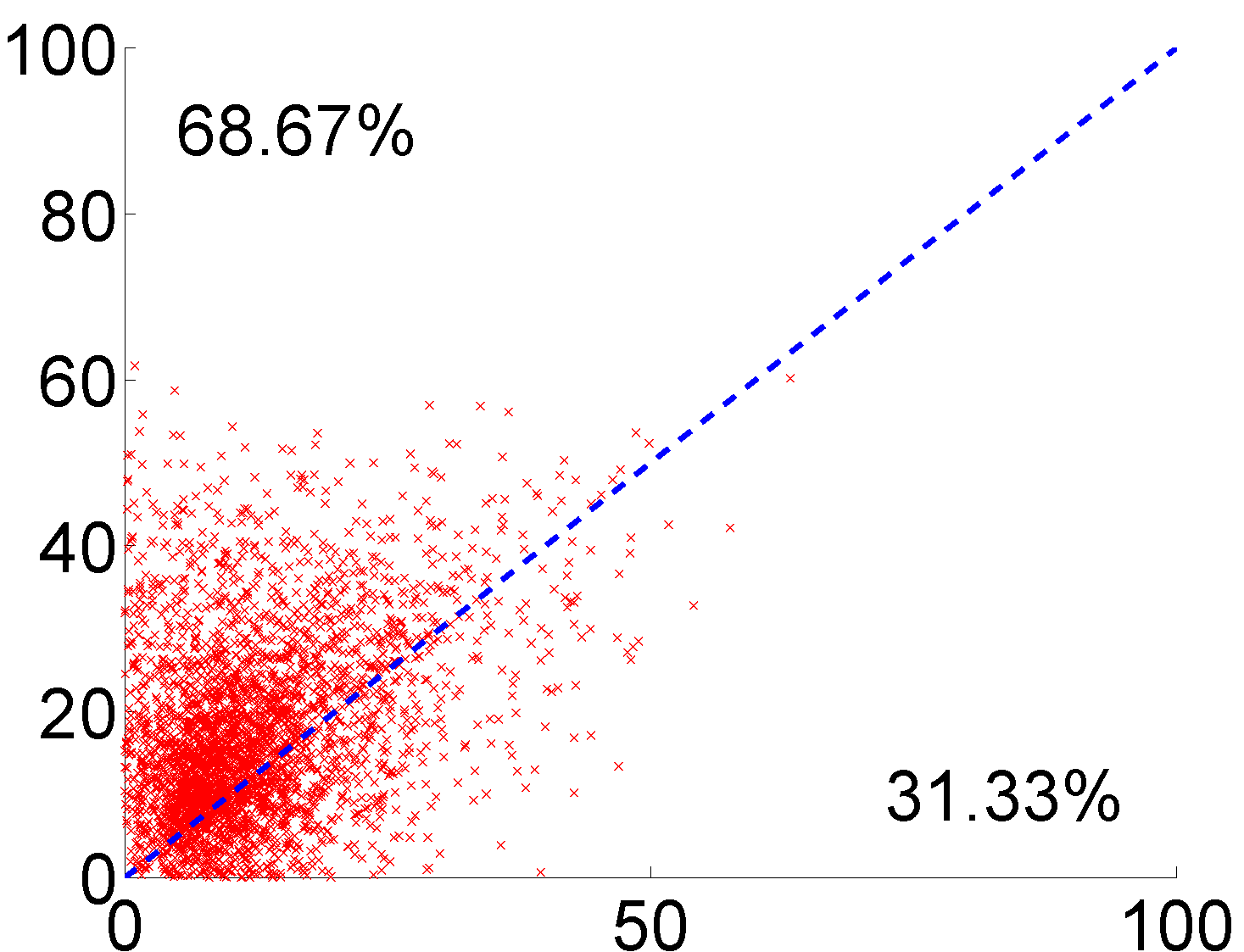} &
\includegraphics[width=1.5in]{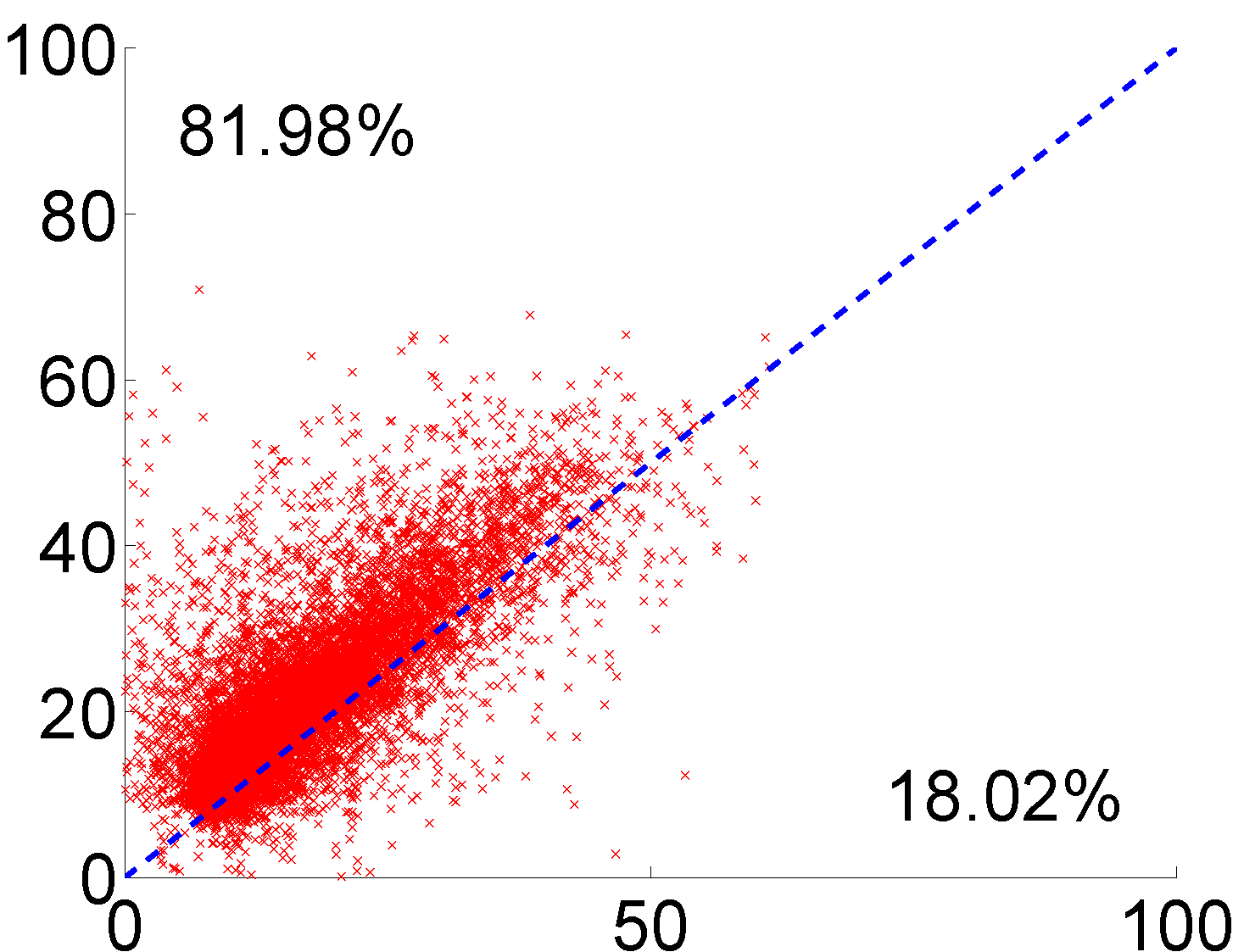} &
\includegraphics[width=1.5in]{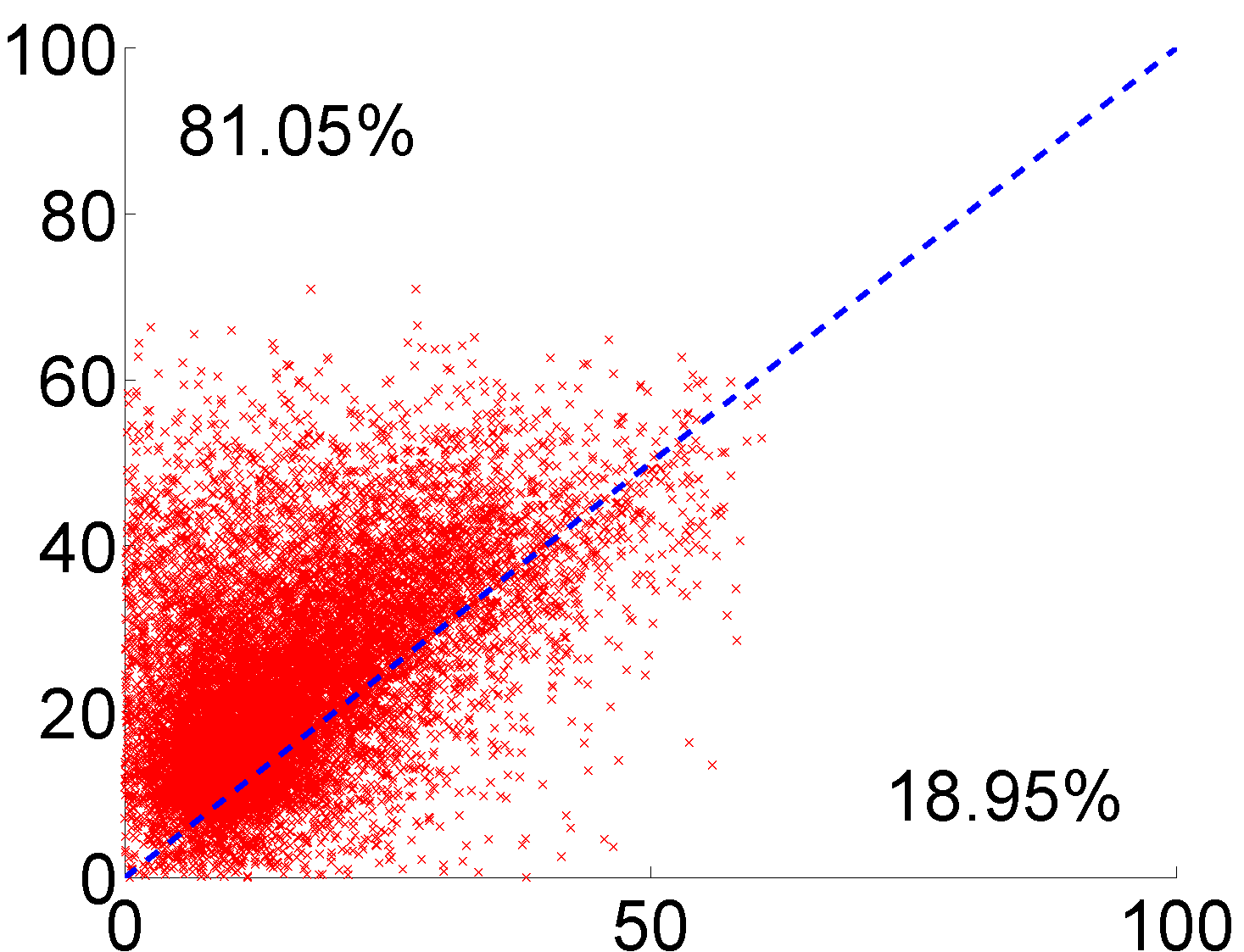}\\
{(a) \#android} &
{(b) \#ladygaga} &
{(c) \#marketing} &
{(d) \#sopa}\\
\end{tabular}
\caption{Distribution of $\Delta_{re}$ and $\Delta_{nre}$. $x$-axis indicates $\Delta_{re}$ and $y$-axis indicates $\Delta_{nre}$.}
\label{chap4:fig:homophily}
\end{center}
\end{figure*}

We calculate $\Delta_{re}$ and $\Delta_{nre}$ for all users in dataset except
for those with empty $N_{re}$ or $N_{nre}$. Homophily exists if most users have
$\Delta_{re} < \Delta_{nre}$, indicating that users with similar influence tend
to follow each other. Distributions of $\Delta_{re}$ and $\Delta_{nre}$ in
Figure~\ref{chap4:fig:homophily} show that the hypothesis is generally true on
Twitter. We notice that the percentage of users satisfying $\Delta_{re} <
\Delta_{nre}$ varies from one community to another. Some communities
(\#marketing, \#sopa) have stronger homophily than others (\#android,
\#ladygaga). This is probably because users have more awareness of their
influence in the \#market and \#sopa communities and  tend to befriend those
who have similar influence. There are many factors that affect the following
relationship among users. Our result shows that influence or its social
perception may attribute to the following relationship, which may in turn
affects a user's influence score.


\section{First-Influencer Information Diffusion Model}
\label{sec:fimodel}
To this end, we have investigated the follower-following relations between
users in Twitter communities, and their connections to influence scores. In
this section, we answer the question of how information spreads in a Twitter
community and propose the FI diffusion model, which is shown empirically to be
more accurate than the widely adopted IC model.

\subsection{Motivation}

Diffusion models that explain how information is spread or how a product is adopted can generally be divided into two categories:
(1) {\it Threshold models}~\cite{ThresholdModel} where each node has a random
threshold and will be activated if the cumulative influence from its neighbours
is larger than its threshold. (2) {\it Cascade models}~\cite{CascadeModel}
where each node when first becomes active will have a chance to activate each of its
inactive neighbours in the next time slot with a certain (edge) probability.
%

Among them, the IC model~\cite{ICmodel}, where the influence probability from
user $u$ to user $v$, $p_{u,v}$ is a constant, has been widely adopted in
literature. Many work~\cite{Singer2012,Bakshy2011,Meeder2011} applied the IC
model in solving the influence maximization problem or studying the spread of
information on Twitter. However, we observe that the current Twitter
implementation does not support such a spreading mechanism. When a user $u$
tweets a new message $m$, this message will be visible to his followers. In
other words, $u$ attempts to spread $m$ to all of his followers.  We say that
$m$ is spread from $u$ to one of his followers $v$ if $v$ retweets $m$.  The IC
model assumes that if the first spread attempt fails, later attempts can still
succeed with constant probabilities.  As an example, $u$ can spread $m$ to $v$
with probability of success $p_{u,v}$.  Assuming $u$ fails, but $m$ was later
spread to $u'$ who is a followee of $v$, then $m$ will have another chance to
be retweeted by $v$ with probability $p_{u',v}$. We notice that unlike other
online social networks such as Facebook, the current implementation of Twitter
(both web and mobile platforms) suppresses the duplicated message.  In other
words, $v$ will not be aware of the fact that $u'$ retweets $m$ and thus, $m$
has no chance to be retweeted by $v$ (or $u'$ cannot influence $v$) if it fails
in the first try.



This observation motivates us to propose a new influence diffusion model -- FI,
to capture the effect that only the first followee who attempts to spread the
influence counts.

\subsection{The model}
Consider a Twitter community modelled by the network $G=\{V,E\}$, with $V$ and
$E$ are the sets of nodes and edges respectively. An active node $u$ at time
$t$ will attempt to spread the information $a$ to one of its inactive neighbours
$v$ with probability of success $p^{a}_{u,v}$, given that $v$ is yet to be
activated by any other nodes. If $v$ is activated, it will in turn, try to
spread the information and influence its inactive neighbours in time $t + 1$. However, if $v$ fails to be
influenced at the first attempt, it will show strong resistance to similar
attempts from it neighbours in the future and set $p^{a}_{u',v} = \varepsilon$
for all $u' \in N_{in}(v)$, where $\varepsilon$ is a small number. In this
work, we consider the special case where $\varepsilon = 0$. 

Formally, we consider three types of sets: $A(t)$, $U(t)$ and $S(t)$, respectively, for
the set of active nodes, the set of insusceptible nodes, and the set of
susceptible nodes at time $t$. Influence propagates in a slot-by-slot manner.
Initially, $A(0) = A$ is the set of seed nodes, $U(0)= \emptyset$, $S(0) =
V\backslash A(0)$. Let $a(t)$ be the collection of newly activated nodes at the
beginning of slot $t$. A node $u \in a(t)$ selects a timer $d_u$ uniformly
distributed in $[0,1]$. Upon the expiration of the timer, $u$ attempts to
spread the information to its inactive neighbours (in $S$). At the end of slot
$t$, a node $v \in S(t)$ is activated with probability $p_{u,v}$ if $u \in a(t)
\cap N(v)$ and $d_u = \min_{x \in N(v)\cap a(t)}{d_x}$; otherwise, $U(t+1) =
U(t)\cup\{v\}$.  

Let $\sigma(A)$ be the expected number of nodes that the seed set $A$ can
influence on the network (also referred to as the spread function).  An
important question is if $\sigma(A)$ is submodular and monotone under FI as
these properties gives rise to efficient approximation
algorithms~\cite{kempe03}. 

\begin{property}
Under the FI model, $\sigma(A)$ is non-monotone. 
\end{property}
\begin{proof}
Consider a network with three nodes $u_1, u_2, u_3$ with edges between $u_1$
and $u_3$, and $u_2$ and $u_3$. Let the edge influence probability $p_{1, 3} >
p_{2, 3}$. It is easy to show that $\sigma(\{u_1\}) = p_{1,3}$, and
$\sigma(\{u_1, u_2\}) = \frac{p_{1,3} + p_{2,3}}{2}$. The second equation is
because when $A = \{u_1, u_2\}$, $u_1$ and $u_2$ each has 0.5 chance of being
the first influencer. Since $p_{1,3} > p_{2,3}$, $\sigma(\{u_1, u_2\}) <
\sigma(\{u_1\})$.
\end{proof}

\begin{property}
Under the FI model, $\sigma(A)$ is non-submodular. 
\end{property}
\begin{proof}
Consider a network with four nodes $u_1, u_2, u_3, u_4$ with edges from $u_i$
to $u_4$ for $i=1, 2, 3$.  The edge influence probabilities satisfy, $p_{1, 4}
> \frac{p_{2, 4} + p_{3,4}}{2}$.  It is easy to show that $\sigma(\{u_1, u_3\})
- \sigma(\{u_1\}) = \frac{p_{3, 4} - p_{1,4}}{2}$ and $\sigma(\{u_1, u_2,
  u_3\}) - \sigma(\{u_1, u_2\}) = \frac{p_{3,4} - p_{2,4} - p_{1,4}}{3}$.  Due
to the condition that $p_{1, 4} > \frac{p_{2, 4} + p_{3,4}}{2}$, we have
$\sigma(\{u_1, u_3\}) - \sigma(\{u_1\})  < \sigma(\{u_1, u_2, u_3\}) -
\sigma(\{u_1, u_2\})$. Thus, submodularity is violated. 
\end{proof}

From the claims, we can see that the FI model differs from the decreasing
cascade model previously proposed by Kempe {\it et al.} that have been proven
to be submodular and monotone~\cite{CascadeModel}.  One key difference is that
in the decreasing cascade model, the activation probability of a node is
independent of the order of nodes that try to influence it, while in the FI
model, the one that first tries to influence the interested node always
dominates as evident from the Twitter implementation~\footnote{The equivalence
between the decreasing cascade model and the generalized threshold model in
\cite{CascadeModel} holds only when the later is non-monotone. In other words,
though we can represent the FI model using the generalized threshold model,
there is no corresponding decreasing cascade model.}.  Despite the above
negative results, we show empirically in the subsequent section that the FI
model is more accurate than the IC model in the prediction of influence spread
in Twitter networks. In both IC and FI models,  the influence spread function
of any given seed set can be evaluated via Monte Carlo simulations. 

\subsection{Evaluation}
In this section, we conduct experiments to compare the proposed FI model with
the IC model.  Since there is no ground truth of the actual diffusion model, we
present two sets of evaluation study on Twitter datasets to demonstrate the
advantages of the proposed model. Experiments are carried out on a variety of
communities \#android, \#at\&t, \#family guy, \#hiphop, \#iphone, \#teaparty to
evaluate the effects of different network structures and densities.

\paragraph*{\bf Model stability:}
In the first set of experiments, we assess the stability of each model.  If a
model is a suitable one, it should have similar parameters for similar datasets
(though the reverse is not necessarily true).

We first extract information cascades from the datasets using the algorithm
in~\cite{goyalinfluencelearning} and put them in a log, referred to as the {\it
cascade log}. In each round of the experiments, we randomly shuffle the
cascades in the {\it cascade log} and break them into 2 equal sets. A model is
considered more stable if the parameters derived from the two sets are more
comparable. A similar metric was adopted in \cite{goyalinfluencelearning}. The
parameters that we infer are the influence probabilities on the edges. By
applying the algorithm in~\cite{goyalinfluencelearning} on both sets, we
calculate the probability vectors $\pp_1, \pp_2 = [p_{u,v}]_{1\times m}$ with
any edge $(u,v)\in E$, respectively, where $m = |E|$ from the two sets.  The Root
Mean Square Error between $\pp_1$ and $\pp_2$ is derived as $RMSE(\pp_1,\pp_2)
= \sqrt{\frac{\sum_{i=1}^{m}{(p_1(i) - p_2(i))^2}}{m}}$.  Denote by $RMSE_{FI}$
and $RMSE_{IC}$ the value of $RMSE$ calculated from the FI and IC models
respectively. $RMSE$ indicates how  much deviation the two set of inferred
parameters exhibit. Higher $RMSE$ implies the model is less stable.  We conduct
100 rounds of experiments and report the average RMSE and the standard
deviation.


\begin{figure*}[t]
\begin{center}
\begin{tabular}{ccc}
\includegraphics[width=2.0in]{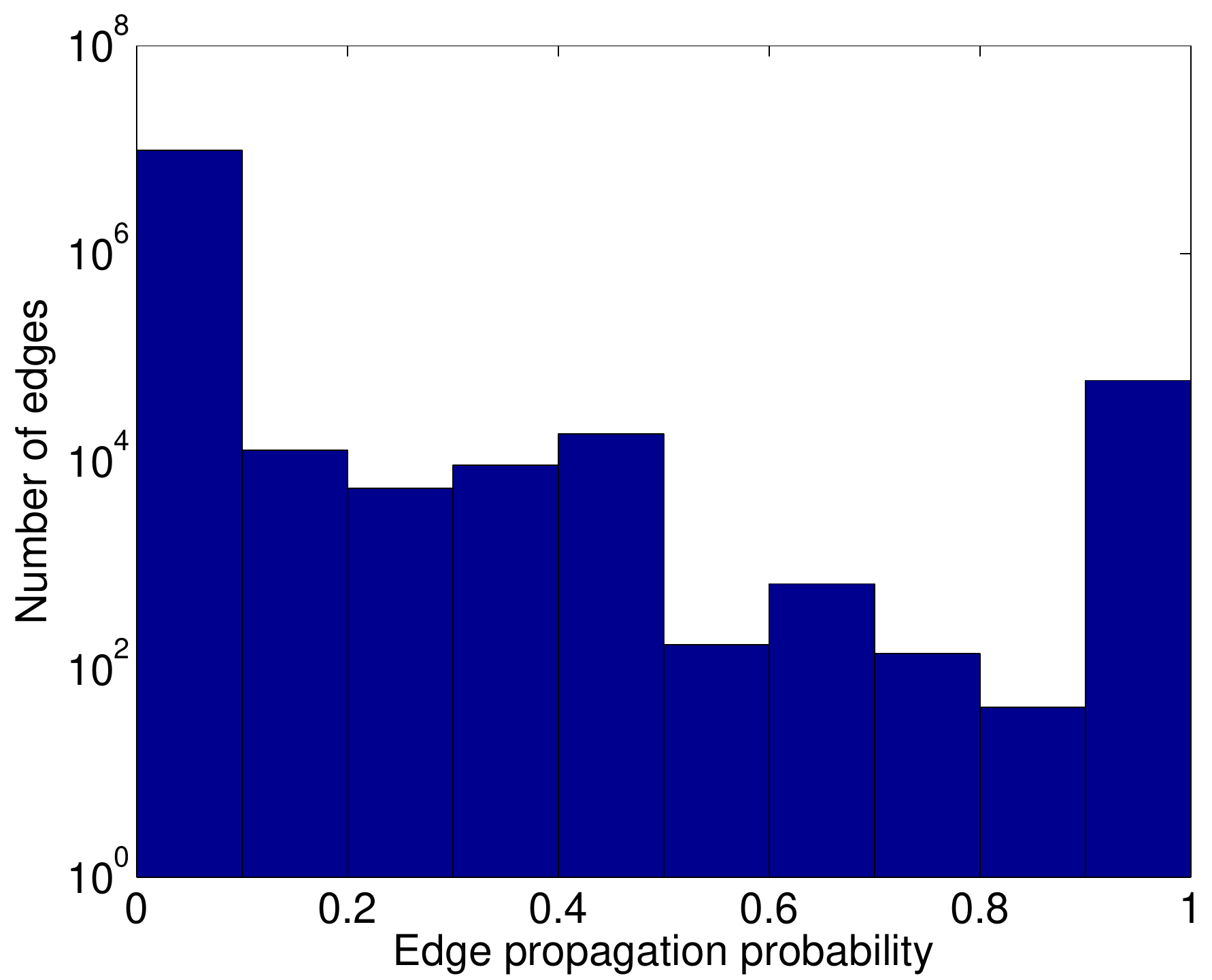} &
\includegraphics[width=2.0in]{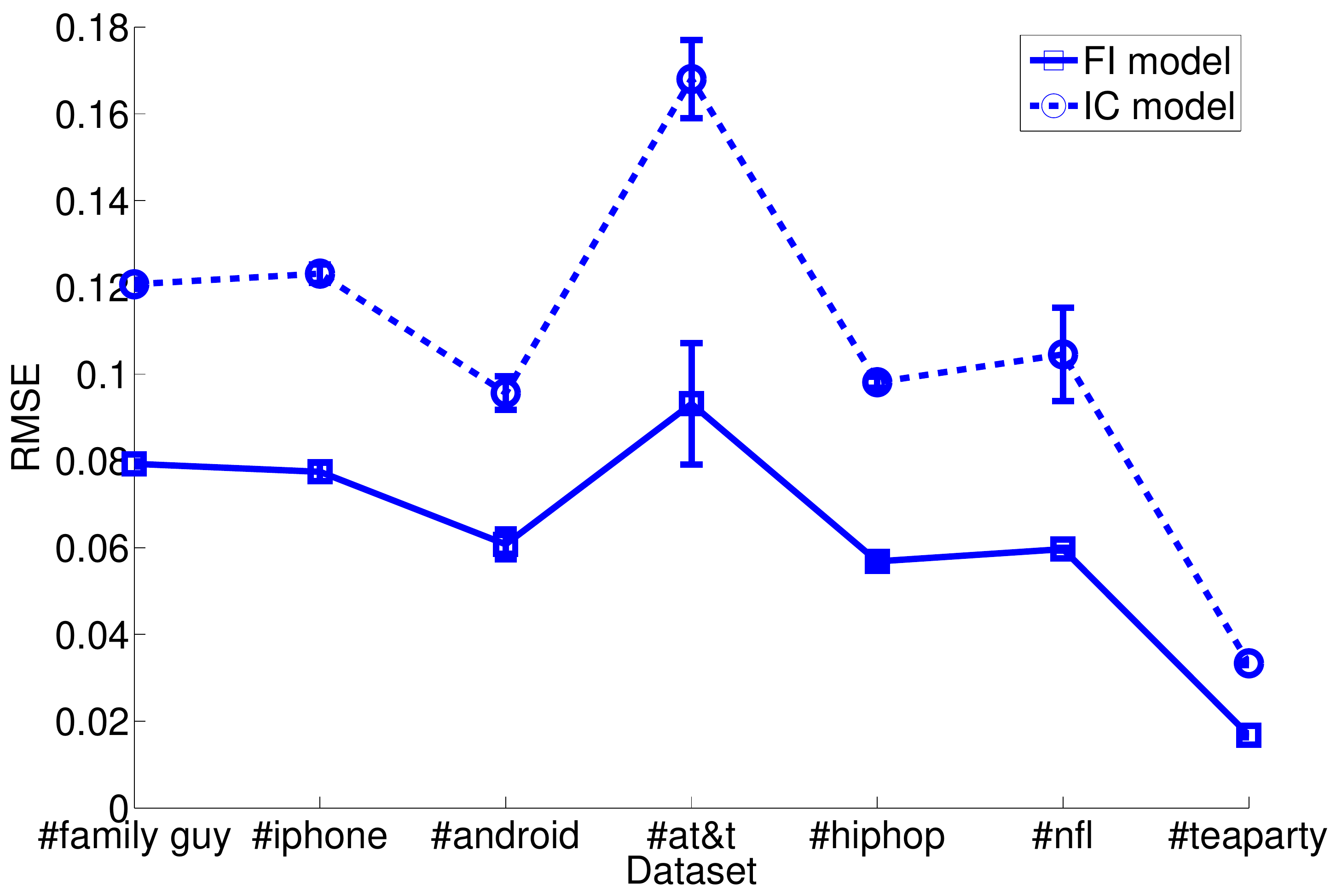} &
\includegraphics[width=2.0in]{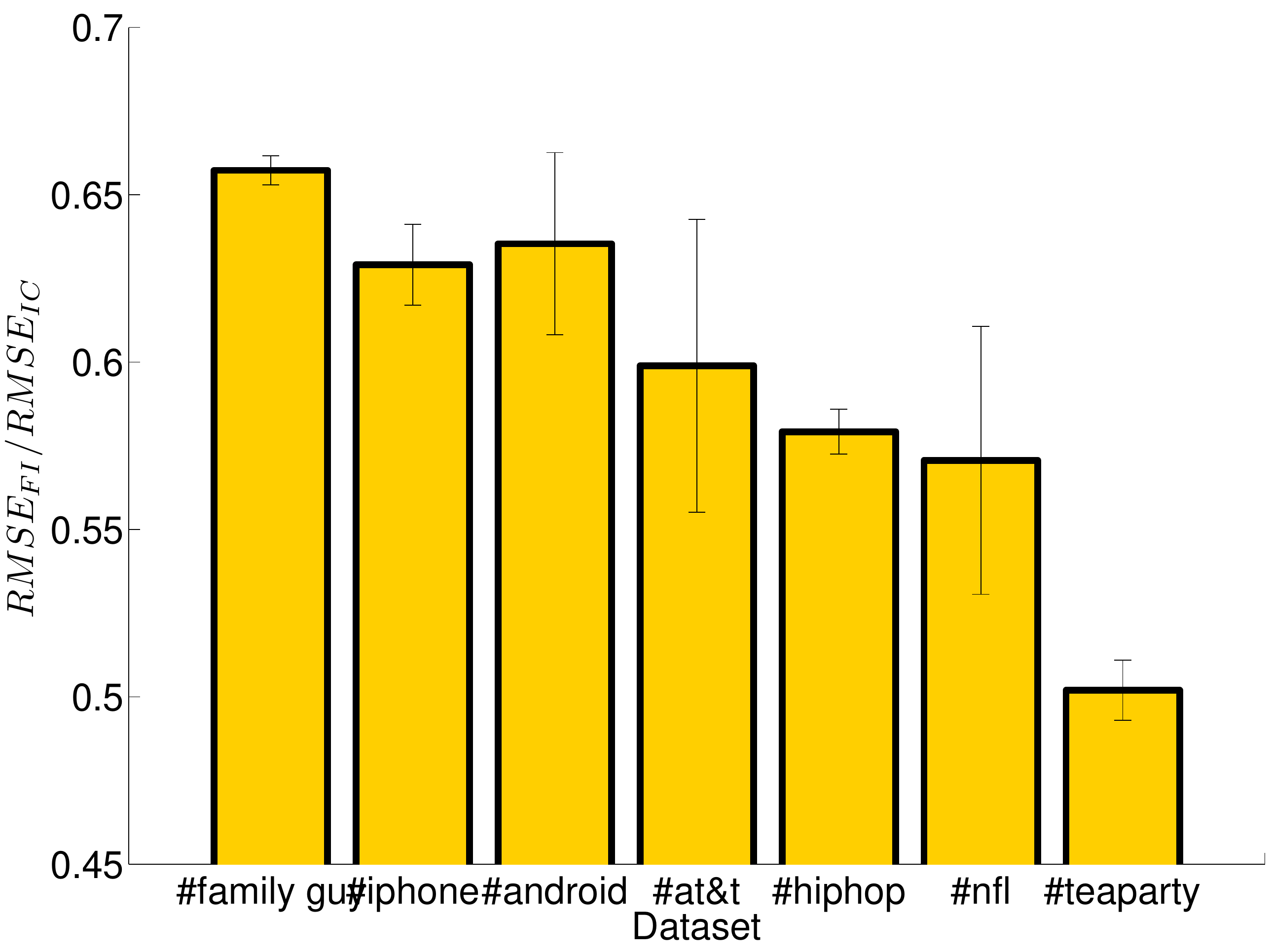}\\
{(a) Histogram of inferred} &
{(b) $RMSE$ on 7 datasets} &
{(c) $RMSE_{FI} / RMSE_{IC}$ ratio}\\
{probabilities on 7 datasets} & &

\end{tabular}
\caption{\small Model stability comparison. Error bars indicate standard deviation.}
\label{chap4:fig:modelstability}
\end{center}
\end{figure*}

Figure~\ref{chap4:fig:modelstability}(a) shows the histogram inferred $\pp_1$
and $\pp_2$ on 7 datasets in log scale. We observe that though predominately
the edge influence probabilities are small, there are many edges having
influence probabilities close to 1. This is due to the small number of cascades
in the datasets along some edges. For example, if there is only one message
from $A$ to $B$ and $B$ retweets, the edge influence probability from $A$ to
$B$ is 1.  From Figure~\ref{chap4:fig:modelstability}(b), we see that both
$RMSE_{FI}$ and $RMSE_{IC}$ are small in all communities. However $RMSE_{FI}$
is consistently smaller than $RMSE_{IC}$, demonstrate the superior stability of
the FI model.  Figure~\ref{chap4:fig:modelstability}(c) gives the ratio between
$RMSE_{FI}$ and $RMSE_{IC}$. The $x$-axis corresponds to communities with
increasing densities from left to right. On dense networks, the IC model tends
to overestimate the spread probabilities since each active neighbour has a
chance to influence. As a result, we see a larger performance gap between FI
and IC when the network is denser (e.g., the tea party community).


\paragraph*{\bf Influence spread prediction:}
The second set of experiments aim to evaluate which model is more accurate in
predicting the influence spread in the network. Influence spread is defined as
the average number of users influenced. First, we derive the edge influence
probability using the method discussed before.  However, in this set of
experiments, the propagation probability is determined from the whole {\it
cascade log}. Denote by $\pp_{FI}$ and $\pp_{IC}$ the probability vectors from
the FI and IC models, respectively. We compute the vectors of expected spread
$\sigma_{FI}, \sigma_{IC} = [\sigma(u)]_{1\times n}$ from $\pp_{FI}$ and
$\pp_{IC}$ for each user $u \in V$ and $n = |V|$ using 10,000 rounds of Monte
Carlo simulations (since the exact calculation of $\sigma$ from $\pp$ is
\#P-complete~\cite{kempe03}). To obtain the ground truth $\sigma(u)$, we
calculate the average size of cascades from a user $u$.
%

\begin{figure}[t]
\begin{tabular}{p{3.5cm}p{3cm}}
\includegraphics[width=1.4in]{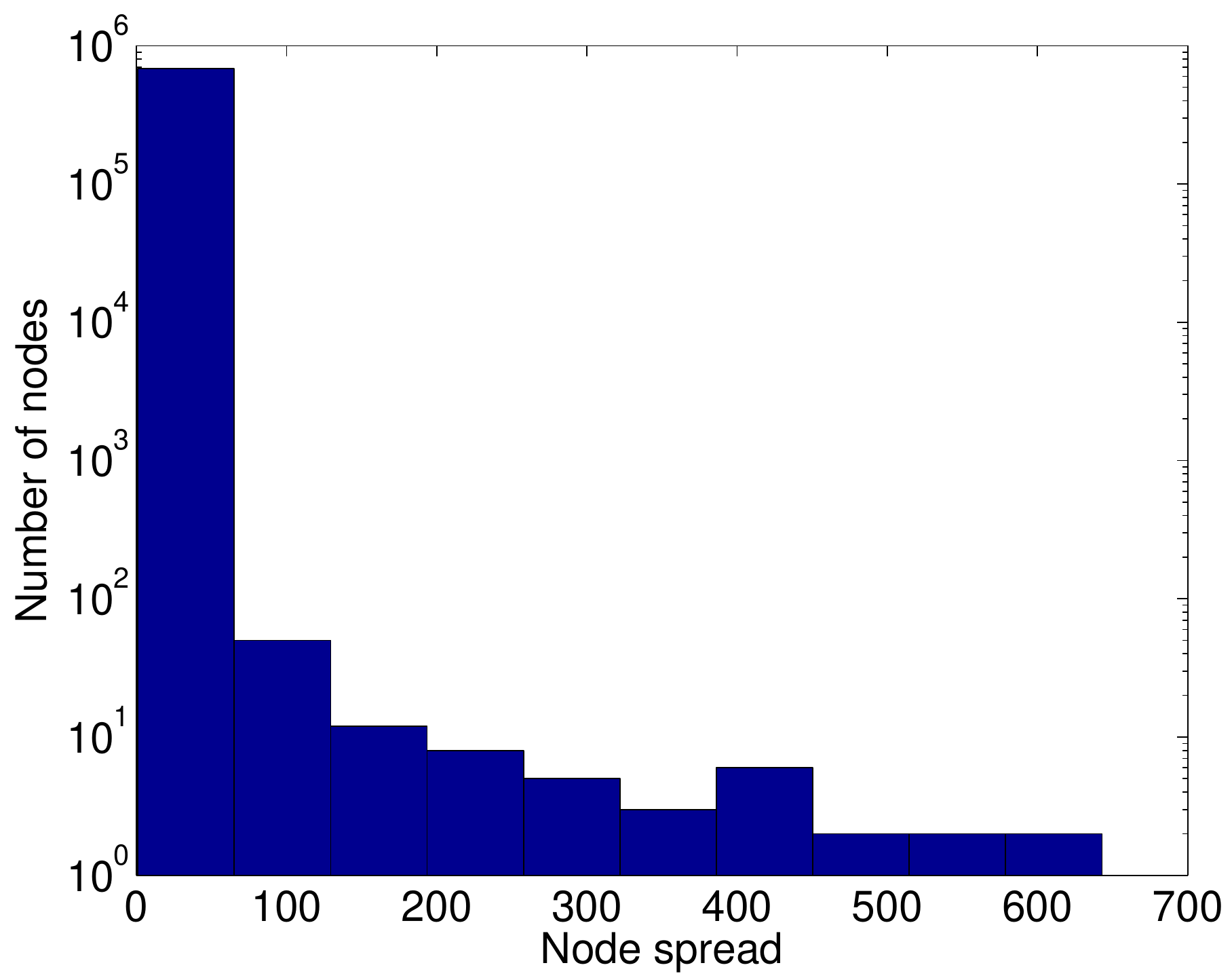} &
\includegraphics[width=1.6in]{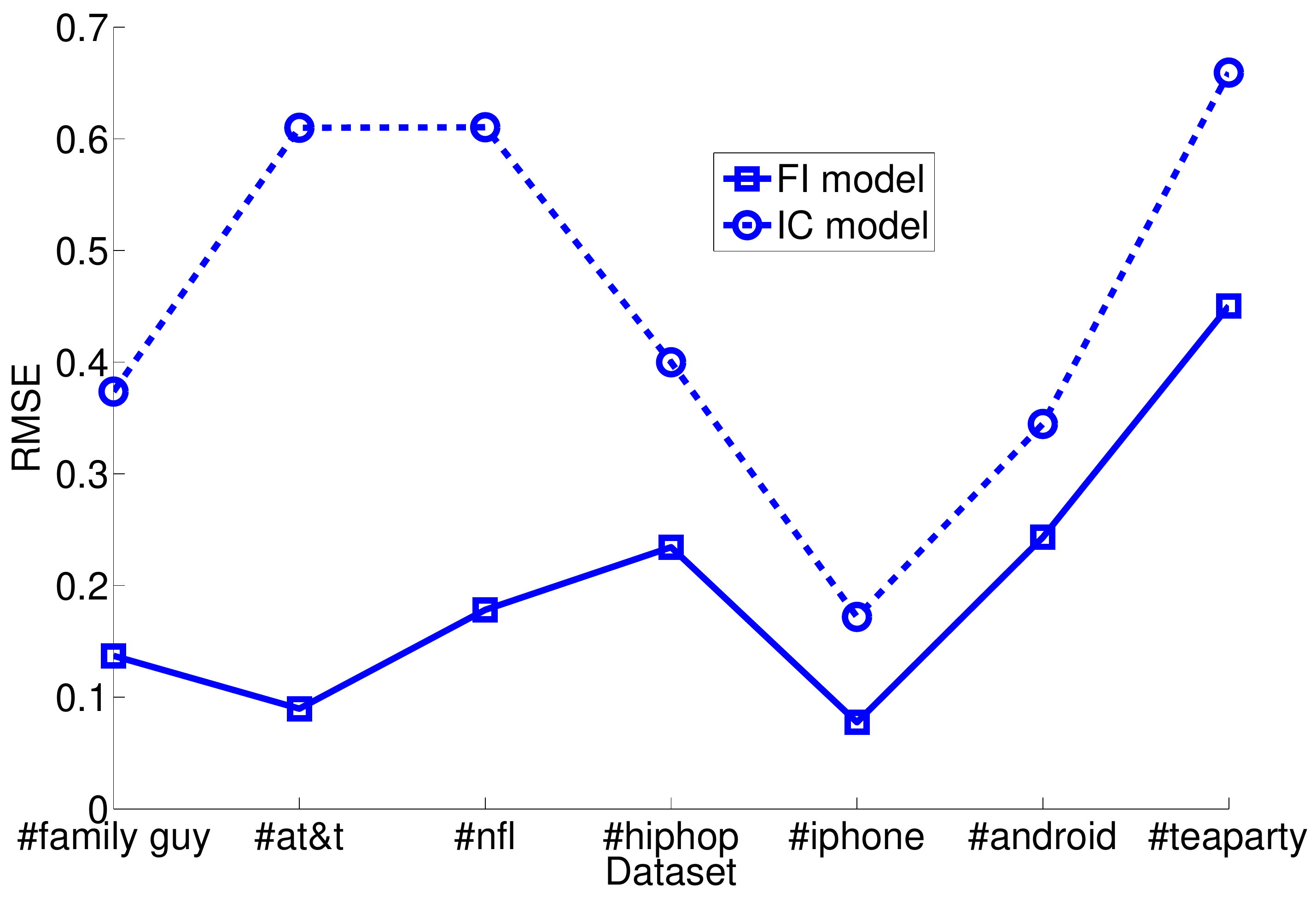}\\
{\small (a) Histogram of $\sigma$ on 7 datasets} &
{\small (b) Value of $RMSE$ on 7 datasets}
\end{tabular}
\caption{\small Comparison of the influence spread prediction of the FI and IC models. The communities are ordered by increasing densities from left to right.}
\label{chap4:fig:spreadprediction}
\end{figure}

The histogram of $\sigma(u)$ on 7 communities is plotted in
Figure~\ref{chap4:fig:spreadprediction}(a). The majority of users have $\sigma
= 1$, which means they fail to spread the information to any of their
followers. The mean value of $\sigma$ is 1.135. We also calculate the RMSE
between the two distributions of $\sigma_{FI}$ and $\sigma_{IC}$ versus the
ground truth $\sigma$ and present the result in
Figure~\ref{chap4:fig:spreadprediction}(b).  We observe that the FI model
consistently outperforms the IC model and can achieve a more accurate influence
spread prediction in all communities.

\subsection{Discussion}
\label{sec:discussion}
We show through the two set of experiments that the FI model is more stable and
results in more accurate influence prediction than the IC model in the Twitter
communities. Thus, in predicting influence spread on Twitter networks, the FI
model is more appropriate. It is motivated by the current implementation of
Twitter in suppressing duplicated messages, and is thus application specific.
It should be noted that FI may not be suitable for other online social
networks, such as Facebook, Google+, etc. However, we believe our study points
to an interesting direction in devising practical influence propagation models
by examining the actual implementation of messaging mechanisms in these
networks.

Twitter's decision to surpress dupliciated message has implications on the
extent of influence propogation in Twitter communities as evident from the
non-monotonicity of the FI model. 



\section{Conclusions}
\label{sec:conclusion}

In this paper, we made two contributions in characterizing influences in
Twitter communities. First, we conducted a quantitative analysis on Twitter
from the user influence perspective. We found that users who share similar
interests, or have similar influences, are likely to befriend each other. We
also found that twitterers tend to follow those with more influence. Second, we
observed the information diffusion on Twitter and proposed the FI model to
capture the spreading process. The findings provide a more comprehensive
understanding of Twitter characteristics, which has implications in many
application domains, such as viral marketing and recommendation
systems.

\bibliographystyle{acm}
\bibliography{ref_splncs}

\end{document}